\documentclass[journal, twocolumn, 10pt]{IEEEtran}

\usepackage{color}
\usepackage{url}
\usepackage{soul}
\usepackage{balance}
\usepackage{enumitem}

\usepackage[binary-units = true, range-phrase=--, per-mode=symbol]{siunitx}
\DeclareSIUnit{\decibelm}{dBm}
\DeclareSIUnit{\Joule}{Joule}

\ifCLASSINFOpdf
  \usepackage[pdftex]{graphicx}
\else 
\fi

\usepackage{nicefrac}
\usepackage{amsmath, amssymb, amsthm}
\usepackage{mathtools}

\usepackage{algorithm}
\usepackage{algorithmic}

\usepackage{booktabs}
\usepackage[caption=false]{subfig}

\usepackage{url}
\newcommand*{\QEDA}{\hfill\ensuremath{\blacksquare}}%

\newcommand{\fakeparagraph}[1]{\vspace{.5mm}\noindent\textbf{#1.}}
\newcommand{\fakepar}[1]{\fakeparagraph{#1}}
\begin{document}

\title{Energy-Reliability Aware Link Optimization for Battery-Powered IoT Devices with \\Non-Ideal Power Amplifiers}

\author{Aamir~Mahmood,~\IEEEmembership{Member,~IEEE},  
				M.~M.~Aftab~Hossain, Cicek~Cavdar,\\~and~Mikael~Gidlund,~\IEEEmembership{Senior Member,~IEEE}% <-this % stops a space
\thanks{A. Mahmood and M. Gidlund are with the Department of Information Systems and Technology, Mid Sweden University, 851 70 Sundsvall, Sweden, e-mail: aamir.mahmood@miun.se.}% <-this % stops a space
\thanks{M. M. A. Hossain and C. Cavdar are with the Wireless@KTH, KTH Royal Institute of Technology, Sweden.}% <-this % stops a space
\thanks{An earlier version \cite{MahmoodWCNC18} of this paper was presented at the IEEE WCNC Conference and was published in its Proc. DOI:\protect\url{10.1109/WCNC.2018.8377180}}
\vspace{-20pt}
}

\maketitle

\begin{abstract}

In this paper, we study cross-layer optimization of low-power wireless 
links for reliability-aware applications while considering both the 
constraints and the non-ideal characteristics of the hardware in 
Internet of things (IoT) devices. 
Specifically, we define an energy consumption (EC) model that captures the 
energy cost---of transceiver circuitry, power amplifier, packet error 
statistics, packet overhead, etc.---in delivering a useful data bit. We 
derive the EC models for an ideal and two realistic non-linear 
power amplifier models. To incorporate packet error statistics, we develop a 
simple, in the form of elementary functions, and accurate closed-form 
packet error rate (PER) approximation in Rayleigh block-fading. 
Using the EC models, we derive energy-optimal yet reliability and hardware 
compliant conditions for limiting unconstrained optimal signal-to-noise ratio 
(SNR), and payload size. Together with these conditions, we develop a 
semi-analytic algorithm for resource-constrained IoT devices to jointly optimize 
parameters on physical (modulation size, SNR) and medium access control (payload size and 
the number of retransmissions) layers in relation to link distance. Our results 
show that despite reliability constraints, the 
common notion---higher-order $M$-ary modulations (MQAM) are energy 
optimal for short-range communication---prevails, and can provide up to 
\SI{180}{\percent} lifetime extension as compared to often used OQPSK 
modulation in IoT devices. 
However, the reliability constraints reduce both their range and the 
energy efficiency, while non-ideal traditional PA reduces the range further by 
\SI{50}{\percent} and diminishes the energy gains unless a better PA is 
used.   

\end{abstract}


\IEEEpeerreviewmaketitle

%%%%%%%%%%%%%%%%%%%%%%%%%%%%%%%%%%%%%%%%%%%%%%%%%%%%%%%%%%%%%%%%%%%%%%%%%%%
%%%%%%%%%%%%%%%%%%%%%%%%%%%%%%%%%%%%%%%%%%%%%%%%%%%%%%%%%%%%%%%%%%%%%%%%%%%
%%%%%%%%%%%%%%%%%%%%%%%%%%%%%%%%%%%%%%%%%%%%%%%%%%%%%%%%%%%%%%%%%%%%%%%%%%%
%%%%% Introduction
%%%%%%%%%%%%%%%%%%%%%%%%%%%%%%%%%%%%%%%%%%%%%%%%%%%%%%%%%%%%%%%%%%%%%%%%%%%
%%%%%%%%%%%%%%%%%%%%%%%%%%%%%%%%%%%%%%%%%%%%%%%%%%%%%%%%%%%%%%%%%%%%%%%%%%%
%%%%%%%%%%%%%%%%%%%%%%%%%%%%%%%%%%%%%%%%%%%%%%%%%%%%%%%%%%%%%%%%%%%%%%%%%%%
\section{Introduction} 
\IEEEPARstart{I}{n} rapidly growing Internet-of-things (IoT), a broad-range 
of applications---from time-critical services to connectivity of massive 
autonomous devices~\cite{Hamid_MTC}---rely on energy- and 
hardware-constrained radio devices for local-area machine-to-machine type 
{communications} (MTC). In MTC, the devices exchange short packets at {a} low duty 
cycle~\cite{ShortPackets} with a controller or directly between two devices~\cite{ProbQoS} 
with guaranteed reliability and latency. When the devices are 
battery-powered, the main design concern is to minimize energy consumption in
wireless communication module to prolong the lifetime as much as 
possible~\cite{Power_IoT_SysJ, CL_Sensors}. However, energy- and reliability-aware communication are two
conflicting demands that often require design trade-offs.

%As the devices will be 
%mostly battery operated, energy efficiency also becomes a critical issue.  
%Moreover, this novel traffic type uses short packets which undermine the 
%coding gain achievable from large packets \cite{ShortPackets}\cite{Shannon}. 
%All these factors urge a 
%new look not only into physical layer but also cross-layer design in order to 
%ensure reliability under energy constraints. In this paper, we improve the 
%packet error rate (PER) approximations over block-fading channels in order to 
%have a better control over the parameters that determine the system 
%performance and utilize these insights to optimize cross-layer 
%parameters, e.g., packet length, number of retransmission, modulation scheme.

Reliable transfer of information has a direct impact on the needed energy 
consumption\cite{Tania_Alouini}. This is because the reliability depends on 
the bit error or packet error statistics of the wireless channel, which in 
turn are influenced by the choice of link parameters such as transmit power, 
modulation scheme, packet length, etc. If the packet error probability has to 
be reduced to transfer a packet successfully with a limited number of 
retransmissions, the link parameters must be optimized jointly while 
restraining the energy consumption. Note that transmitting the same packet to 
provide reliability not only translates into a higher transmission cost but 
also causes delay violations. As a result, it is becoming imperative to look 
not only into physical (PHY) and {medium access control (MAC)} layers separately, but 
across layers with awareness to extreme energy and reliability constraints 
posed by MTC-IoT devices.

In any energy-aware design, the impact of constraints and the characteristics 
of components in the hardware layer cannot be ignored. In an RF device, the transmitter consumes a 
major chunk of energy for RF signal 
generation. This is because the signal generation at the desired 
output power requires amplification via a {power amplifier (PA)}. However, due 
to the limited efficiency and linearity of traditional PAs, typically 
\SIrange{50}{80}{\percent} of energy is consumed by the 
PAs~\cite{PA_Survey, hossain2011impactPA}. Therefore, the low-power {and low-cost} 
IoT devices require a PA-centric joint design of PHY and MAC parameters.
     
In this paper, we study the selection of optimal---energy consumption 
minimizing and reliability aware---link (PHY and MAC) parameters in 
Rayleigh block-fading channels, however, under the often neglected hardware constraints and 
non-ideal PA characteristics of the IoT devices. To such {an} end, we analyze the {interplay} among 
optimal parameters under a widely used hypothetical {constant PA (CPA)} model, and two 
realistic non-linear models: prevailing traditional PA (TPA), and envelop-tracking 
PA (ETPA)~\cite{hossain2011impactPA, PA_Riku}, which to our best knowledge is the first such study. 

\subsection{Related Works and Contributions}
\label{subsec: RelatedWorks}

To minimize energy consumption in low-power wireless sensor networks (WSNs), 
cross-layer optimization of PHY and MAC parameters has been investigated by 
many authors both in {additive white Gaussian noise (AWGN)} 
channel \cite{cui2005energy, wang2008, hou2005performance} and fading 
channels~\cite{rosas2012modulation, EM_QAM, wu2014energy, Holland_PHY_WSN}. 
These studies suggest using higher-order 
modulations---{$M$-ary quadrature amplitude 
modulation (MQAM)}---at {short} distances, as opposed to the common notion 
followed in power-limited WSNs, which prefer low-order modulations for their low 
SNR requirement. For instance, {highly proliferated} low-power 
transceivers CC1100 and CC2420 in WSNs employ BPSK and OQPSK modulations, 
respectively.

In \cite{rosas2012modulation, EM_QAM, wu2014energy, Holland_PHY_WSN}, an 
energy consumption (EC) model is formulated that corresponds to energy per 
payload bit transferred without error in fading channels. In 
\cite{rosas2012modulation, wu2014energy}, it is shown that, for each 
modulation scheme, there exist an optimal SNR and packet 
length that minimizes the energy consumption. In 
\cite{rosas2012modulation, EM_QAM}, the optimal SNR is conditioned on the 
maximum transmit power, but this constraint is ignored in 
\cite{wu2014energy}. In \cite{Holland_PHY_WSN}, the energy minimization is 
considered via the outage probability instead of packet error statistics. Importantly, 
%Cross-layer optimization of PHY and MAC parameters to minimize 
%energy consumption in low-power wireless sensor networks (WSNs)
 %has been investigated by many authors. How to select 
%the modulation order and transmit power to attain energy-efficient 
%communication are studied in {additive white Gaussian noise (AWGN)} channel 
%\cite{cui2005energy, wang2008, hou2005performance}, and in fading 
%channels~\cite{rosas2012modulation, EM_QAM, wu2014energy, Holland_PHY_WSN}. 
%These studies suggest using higher-order modulations---{$M$-ary quadrature amplitude modulation (MQAM)}---at smaller 
%distances, as opposed to the common notion followed in WSNs where 
%lower-order modulations are preferred for their low SNR requirement. For 
%instance, {highly proliferated} low-power transceivers CC1100 and CC2420 in 
%%WSNs employ BPSK and OQPSK modulations, respectively. In fading channels, it is shown in \cite{rosas2012modulation, wu2014energy} that there exist an optimal SNR 
%and packet length for each modulation scheme at which the required energy for 
%successful transfer of an information bit is minimized. In 
%\cite{rosas2012modulation, EM_QAM}, the optimal SNR or transmit power is conditioned on the maximum 
%transmit power, but this constraint is ignored in \cite{wu2014energy}. 
all these studies assume the system is delay-tolerant, i.e., no 
restriction on the number of retransmissions is imposed. As a result, the 
optimal link parameters are not bound to satisfy the time-critical 
MTC~\cite{Sisinni_IIoT}, {unless retransmissions are adapted
according to reliability and latency 
{constraints}~\cite{MahmoodWCNC18}.} 
Moreover, PA efficiency is assumed to be constant irrespective of the transmit power. %, which is far from realistic PA characteristics. 
However for MQAM modulations, which carry information in both phase and amplitude, the 
realistic PAs suffer from poor power efficiency {because of} 
their non-constant envelops. Therefore, it is yet to be analyzed how these 
modulations behave at short distances under realistic PAs.

%The low cost and low energy communication systems
%tend to exclude the error correction stage and adopt ARQ
%schemes to accommodate the hardware design requirements
%[15]. To this end, we adopt ARQ protocols to
%reduce the system complexity in order to be compatible with
%low cost communication systems.

For link optimization, a valuable cross-layer 
metric capturing the cost-benefit trade-off, is packet error rate 
(PER)~\cite{E_D_Zappone, rosas2012modulation}. However, an exact analytical expression 
of PER in fading channels is not found in the literature. A generic upper 
bound on average PER in Rayleigh block-fading channels is 
$1-\exp(\omega_0/\textrm{SNR})$, where 
$\omega_0$---\textit{the water-fall threshold}---is defined in terms of an 
integral of the PER function in the AWGN channel. Its approximation based on 
log-domain linear approximation of $\omega_0$ is developed for uncoded 
schemes in \cite{liu2012tput}, and for (un)coded schemes in 
\cite{wu2014energy}. However, the approximation in \cite{liu2012tput} is 
tight in {an} asymptotic regime while the approximation 
parameters in \cite{wu2014energy} are simulation-aided. As a result, 
the integral is evaluated numerically, though not computationally intensive, 
but does not offer insights regarding what parameters determine the system 
performance.

{In our previous study~\cite{MahmoodWCNC18}, we developed an 
EC model for cross-layer optimization that captured the energy cost (of 
transceiver circuitry, 
PA, packet error statistics, packet overhead etc.) in delivering a useful 
data bit under prescribed delay and error performance constraints. In 
addition, we tightly approximated $\omega_0$ to get a PER approximation in 
Rayleigh block-fading, which was accurate than \cite{wu2014energy} and \cite{
liu2012tput} while it maintained an explicit relation with the PHY/MAC layer 
parameters unlike~\cite{wu2014energy}. However in \cite{MahmoodWCNC18}, we assumed hypothetical PA 
characteristics in the EC model, and the ensuing optimal parameters 
did not to reflect the true gain in using higher-order modulations for 
short-range communication. In this present work, we significantly consolidate our earlier study by introducing non-ideal 
characteristics of PAs in the EC model and by analyzing their influence on  
the link design.} A summary of our main contributions follows.

\begin{itemize}
	
	%\item We develop energy consumption (EC) models giving energy cost per reliably 
%transmitted information bit and derive it for a hypothetical and two 
%real-life non-linear power amplifier (PA) models. 

\item {We develop energy consumption (EC) models for two 
realistic non-linear power amplifier (PA) models. We find the optimal SNR and 
payload size analytically for the studied PA models, and find the conditions 
for energy-reliability aware selection of SNR and payload size for a 
power-limited system.} 
	
	%\item We show that the \textit{waterfall threshold} ($\omega_0$) is tightly 
%approximated by the expected value of an asymptotic PER distribution in AWGN, 
%which leads to a PER approximation in Rayleigh block-fading accurate 
%than \cite{wu2014energy} and \cite{liu2012tput} while it maintains explicit 
%relation with the PHY and MAC layer parameters 
%unlike~\cite{wu2014energy}.
	
	%\item We find the closed-form expressions of optimal SNR and payload size for the studied PA models, 
%and find the conditions for energy-reliability aware 
%selection of SNR and payload size for a power-limited system.

	\item We propose a joint optimization algorithm to find {the} optimal link 
parameters as a function of link distance under the prescribed reliability 
constraints. We show that the right selection of parameters can increase 
the link's lifetime up to \SI{180}{\percent} compared to OQPSK modulation.

	\item Our analysis with non-linear PA models gives several new insights: under a 
realistic PA, the energy consumption to operate a link reliably increases; {the} 
traditional PA diminishes the gain in using higher (PAPR) modulations at 
short distances; PAs with better linearity (e.g., envelop-tracking PA) 
can improve this situation, and must be studied for power-limited devices.  

\end{itemize}
%
%Unocoded scheme are considered here without lost of generality as it is 
%straight-forward to extend this analysis to coded schemes.

%
%{The contribution of the paper are\\ 
%i) A tight approximation to 
%the waterfall threshold and the PER in Nakagami-$m$ fading channels is one of 
%the contributions in this paper. \\
%ii) ...\\
%iii)}

The rest of the paper is organized as follows. Section~\ref{sec:SystemModel} 
presents the system model and energy-efficiency metrics. 
Section~\ref{sec:EEModel} develops the EC 
model for various PA models, and drives expression for the associated PER 
function. Section~\ref{sec:EE} performs cross-layer optimization with 
reliability constraints and develops an algorithm for link's parameters 
optimization. Section~\ref{sec:Results} gives the insightful results and 
analysis. Finally, conclusions are drawn in 
Section~\ref{sec:Conclusions}.

%%%%%%%%%%%%%%%%%%%%%%%%%%%%%%%%%%%%%%%%%%%%%%%%%%%%%%%%%%%%%%%%%%%%%%%%%%%
%%%%%%%%%%%%%%%%%%%%%%%%%%%%%%%%%%%%%%%%%%%%%%%%%%%%%%%%%%%%%%%%%%%%%%%%%%%
%%%%%%%%%%%%%%%%%%%%%%%%%%%%%%%%%%%%%%%%%%%%%%%%%%%%%%%%%%%%%%%%%%%%%%%%%%%
%%%%% System Model
%%%%%%%%%%%%%%%%%%%%%%%%%%%%%%%%%%%%%%%%%%%%%%%%%%%%%%%%%%%%%%%%%%%%%%%%%%%
%%%%%%%%%%%%%%%%%%%%%%%%%%%%%%%%%%%%%%%%%%%%%%%%%%%%%%%%%%%%%%%%%%%%%%%%%%%
%%%%%%%%%%%%%%%%%%%%%%%%%%%%%%%%%%%%%%%%%%%%%%%%%%%%%%%%%%%%%%%%%%%%%%%%%%%

\section{System Model}
\label{sec:SystemModel}

\subsection{Communication Model}
\begin{figure}[t]
	\centering
		\includegraphics[width=0.95\linewidth]{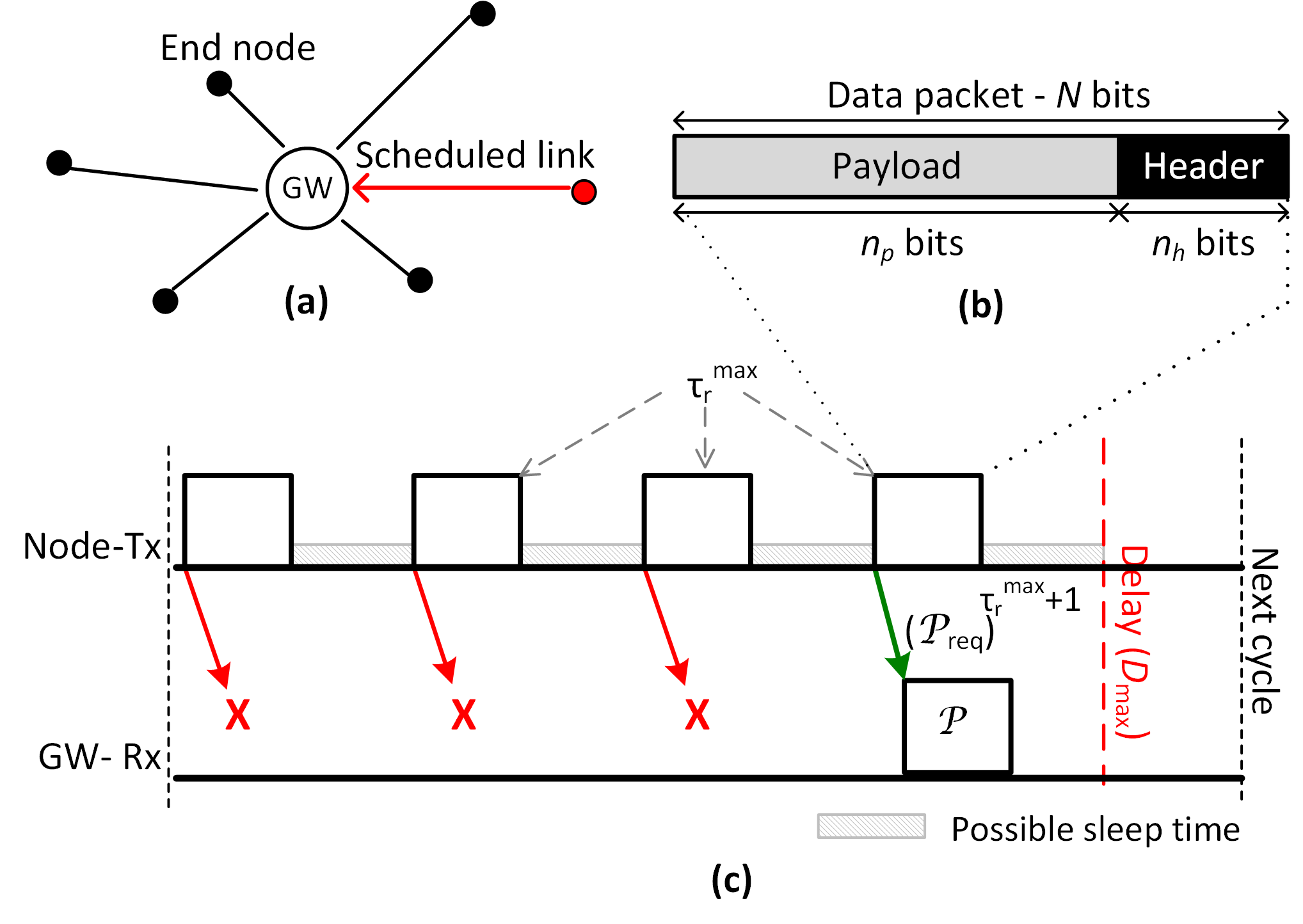}
		\vspace{-8pt}
	\caption{Network model: (a) network 
topology, (b) packet structure,(c) probabilistic service model: after $\tau_r^{\max} + 1$ 
transmissions, a packet must be delivered within delivery deadline ($D_{\max}$) and 
reliability target ($\mathcal{P}$).}
\label{fig:NM}
\vspace{-10pt}
\end{figure}
We consider a master-slave communication model, typical for industrial 
automation applications~\cite{RateAdapt_FA,GW_tut}; a gateway (GW) acting as the 
master, and battery-operated wireless sensors as the slaves (see Fig.~1(a)). The 
sensors are placed at random locations within the communication range of 
the GW. The GW schedules a transmission by broadcasting a short control 
message, which indicates the sensor that shall transmit a packet exclusively 
in time and/or frequency resource. The control message also indicates the 
permissible maximum delay, $D_{\max}$, and the target packet error probability, $\mathcal
{P}$, of the sanctioned transaction (see Fig.~1(c)). Note that the design of a 
scheduling policy that could satisfy the end-to-end latency including the 
delay in scheduling of a sensor is beyond the scope of this paper although  
one may refer to \cite{ProbQoS}.
%---the time from the generation of a packet at a sensor to its reception at 
%the GW---requirment 

%In this model, we make the following assumptions. 
%
%We assume that the control messaging is extremely reliable through the use 
%of, for instance, robust modulation and power control scheme etc.  

\subsection{Metrics, QoS Model and Objectives}

In this work, we are interested in minimizing the energy consumption of the 
wireless sensor devices in transmitting the packet to the gateway. 
While optimizing the energy efficiency, we also want to ensure that the 
{quality-of-service} (QoS) requirements are fulfilled. 

\fakepar{Energy efficiency metric}
To capture the energy consumption and reliability trade-off, the 
cost-benefit of the link is expressed as the ratio of the energy consumption 
to the corresponding data reliably delivered as
\begin{equation}
E = \left(\frac{1}{1-p}\right) E_0,
\label{eq:E}
\end{equation}
where the first term is the average number of transmissions, $\bar{\tau}$, over an i.i.d 
channel realizations between transmissions, which is the mean of a geometric random 
variable with parameter $p$--the packet error rate (PER). The term $E_0$, to be defined precisely in 
Sec.~\ref{sec:EEModel}, accounts for the energy consumed by the radio circuitry to operate the link. The measure unit 
of \eqref{eq:E} is [\si{\Joule\per\bit}] as it represents the amount of 
energy required to transmit a given amount of data or, as stated otherwise, the 
energy cost per reliably transmitted information bit. 

The measure of energy consumption in \eqref{eq:E}, with its relation to PER, 
depends on 
the physical parameters as SNR or transmit power, modulation order and symbol 
rate, and MAC layer parameters including packet size---information and 
overhead bits (see Fig.~1(b))---and the number of retransmissions. 

%Our objective is 
%to maximize EE by minimizing energy consumption per successful transfer of an 
%information bit. 

%The total energy consumption of a wireless link depends on the required 
%retransmissions before a packet is decoded successfully at the receiver. The 
%retransmission statistics are determined by the PER, 
%$\bar{P}_e\left(\bar{\gamma}\right)$, which is a function of $\bar{\gamma}$, 
%channel fading, and other parameters as discussed earlier. The number of 
%retransmissions $\tau$ is geometric random variable and over an uncorrelated 
%channel between retransmissions the average number of retransmissions are $
%\bar{\tau}  = {1}/{(1-\bar{P}_e(\bar{\gamma}))}$. For notational brevity, let 
%us define: $p := P_e\left(\bar{\gamma}\right)$. Therefore, the total average 
%energy for a successful transmission of a bit is $E = \bar{\tau}E_0$, which 
%from \eqref{eq:E_0} is

\fakepar{Energy efficiency metric with probabilistic QoS model}
Since delay $D_{\max}$ is finite, the maximum number of retransmissions has 
to be bounded i.e., $\tau_r^{\max} = \left\lfloor {D_{\max}}/{t_T}
\right\rfloor - 1$ where $t_T$ is the transmission time of the packet. 
Moreover, due to finite retransmissions, an error-free delivery cannot be 
ensured. Since a packet loses its value after $D_{\max}$ in typical 
control applications, the packet is dropped after $\tau_r^{\max}$ 
retransmissions. Therefore, the condition: the probability of packet error 
after $\tau_r^{\max}$ retransmissions is not greater than $\mathcal{P}$, 
defines a probabilistic QoS model~\cite{liu2004cross}
\begin{equation}
p^{\tau_r^{\max} + 1} \leq \mathcal{P}.
\label{eq:PER_t}
\end{equation}
Solving \eqref{eq:PER_t} for $p$ gives the reliability needed at the 
PHY layer under a limited number of retransmissions at the MAC layer 
\begin{equation}
p \leq \mathcal{P}^{1/(\tau_r^{\max} + 1)} := 
\mathcal{P}_{\mathrm{req}}.
\label{eq:PER_req}
\end{equation}   

If \eqref{eq:PER_req} is satisfied for each transmission attempt at the 
physical layer, the application QoS requirement in \eqref{eq:PER_t} with maximum $\tau_r^{\max}$ retransmissions will be 
satisfied with the average number of transmissions per packet 
$\bar{\tau}_{\mathrm{trunc}}  = {1-p^{\tau_r^{\max} + 1}}/{(1-p)}$. In this 
case, the energy consumption follows from 
\eqref{eq:E} as
\begin{equation}
E_{\mathrm{trunc}} = \left(\frac{1-p^{\tau_r^{\max} + 1}}{1-p}\right) E_0.
\label{eq:E_trunc}
\end{equation} 

\fakepar{Objectives} 
From the above-defined probabilistic QoS model, our objective is to find the modulation 
scheme and its operational parameters: SNR at the PHY
layer, and packet size at the MAC layer such that the required PER in 
\eqref{eq:PER_req} is satisfied by implementing $\tau_r^{\max}$ retransmissions at the MAC 
while energy efficiency in \eqref{eq:E_trunc} for the sanctioned 
transaction is maximized.

%In other words, we address the following interesting question:
%with the aid of -truncated ARQ at the data link layer,
%how can we optimally design the AMC at the physical layer
%to maximize throughput, while guaranteeing the overall system
%performance dictated by C2?
%
%Therefore, if we design AMC to satisfy a PER upper-bound 
%as in (3) at the physical layer, and implement a -truncated ARQ at the data 
%link layer, both delay and performance requirements C1 and C2 will be 
%satisfied. Our remaining problem is to design AMC to maximize spectral 
%efficiency while maintaining (3).

%\subsection{Objectives}   
%We consider minimizing energy consumption of a wireless link between a 
%transmitter and receiver pair separated by distance $d$. 

%%%%%%%%%%%%%%%%%%%%%%%%%%%%%%%%%%%%%%%%%%%%%%%%%%%%%%%%%%%%%%%%%%%%%%%%%%%
%%%%%%%%%%%%%%%%%%%%%%%%%%%%%%%%%%%%%%%%%%%%%%%%%%%%%%%%%%%%%%%%%%%%%%%%%%%
%%%%%%%%%%%%%%%%%%%%%%%%%%%%%%%%%%%%%%%%%%%%%%%%%%%%%%%%%%%%%%%%%%%%%%%%%%%
%%%%% Elements of Energy Consumption Model
%%%%%%%%%%%%%%%%%%%%%%%%%%%%%%%%%%%%%%%%%%%%%%%%%%%%%%%%%%%%%%%%%%%%%%%%%%%
%%%%%%%%%%%%%%%%%%%%%%%%%%%%%%%%%%%%%%%%%%%%%%%%%%%%%%%%%%%%%%%%%%%%%%%%%%%
%%%%%%%%%%%%%%%%%%%%%%%%%%%%%%%%%%%%%%%%%%%%%%%%%%%%%%%%%%%%%%%%%%%%%%%%%%%

\section{Elements of Energy Consumption Model}
\label{sec:EEModel}
In this section, we define energy consumption model $E_0$, and derive it for 
various power amplifier models. Also, we find a PER expression that 
maintains an explicit connection with the modulation order and the associated 
parameters. 

\subsection{Average Energy Per Information Bit}
\label{subsec:E_0_Model}
The energy consumption of the signal path between the transmitter and receiver is 
{composed of} baseband processing blocks and radio-frequency (RF) chain. The RF chain consists of a power amplifier (PA) and 
other electronic components such as analog-to-digital and digital-to-analog converters, low-noise amplifier, filters, mixers and frequency 
synthesizers. However, for an energy-constrained wireless system,  
the energy consumption of RF chain is the orders of magnitude higher than the consumption of 
baseband processing blocks. The power consumption of PA is considered to be proportional to the transmit power $P_t$ as $P_{\mathrm{PA}} = \xi P
_t/\eta(P_t)$, where $\xi$ is the modulation scheme dependent peak-to-average power 
ratio (PAPR) and $\eta (P_t)$ is the $P_t$-dependent drain efficiency of the PA (see Sec.~\ref{subsec:realisticPA}). If baseband power 
consumption is neglected and the power consumption of all the other 
components in RF chain excluding PA is denoted as $P_c$, a simple power 
consumption model is $P_{\mathrm{on}} = \xi P_t/\eta(P_t) + P_c$. From 
\cite{cui2005energy}, it leads to total energy consumption to 
transmit and receive a symbol, $E_{\mathrm{sym}}$
\begin{equation}
E_{\mathrm{sym}} = \frac{\xi E_t}{\eta(P_t)} + \frac{P_c}{R_s},
\label{eq:E_symbol}
\end{equation}
where $E_t$ is the average transmission energy of a symbol and $R_s$ is the 
symbol rate at PHY. For FSK, BPSK and QPSK modulations, PAPR $\xi = 1$, for OQPSK $\xi = 2.138$, and for a 
square MQAM modulations, it can be approximated as $\xi = 3 (\sqrt{M} - \frac{1}{\sqrt{M}} + 1)$
\cite{cui2005energy}.

Let $E_b = {E_r}/{\log_2M}$ be the average received energy per uncoded 
bit, where $E_r$ is the average received energy per symbol and $M$ is the 
constellation size. Then, the average SNR ($\bar{\gamma}$) at the receiver is 
\begin{equation}
\bar{\gamma} = \frac{E_r}{N_0\log_2M}.
\label{eq:snr}
\end{equation}

Assuming a $\kappa$th-power path-loss model, the transmission energy at 
distance $d$ from \eqref{eq:snr} can be expressed as \cite{cui2005energy}
\begin{equation}
E_t \triangleq E_r G_d = \Big(\bar{\gamma}N_0 \log_2 M\Big) G_d,
\label{eq:E_t}
\end{equation}
where $G_d \triangleq G_1 d^\kappa M_\ell$ is the path-loss gain, where $G_1$ is
the gain factor at unit distance, which depends on the transmit and 
receive antenna gains and carrier frequency, and $M_l$ is the link margin.

In a {packet-based} wireless system, the information bits are encapsulated into 
packets each carrying $n_p$ payload and $n_h$ overhead bits; thus the number 
of symbols in each packet are $n_s = (n_h + n_p)/\log_2 M$. Therefore, the 
average energy required to transmit and receive an information bit, from 
\eqref{eq:E_symbol} and \eqref{eq:E_t}, is
\begin{equation}
E_{0} = \left(\frac{n_p + n_h}{n_p \log_2 M}\right) E_{\mathrm{sym}}.  
\label{eq:E_0}
\end{equation}

%%%%%%%%%%%%%%%%%%%%%%%%%%%%%%%%%%%%%%%%%%%%%%%%%%%%%%%%%%%%%%%%%%%%%%%%%%%%%%%%
\subsection{$E_0$ for Different Power Amplifier Models}
\label{subsec:realisticPA}

As PA is the most power consuming component of a wireless link, its real-life 
characteristics must be considered for energy efficient communication. The 
power consumption of PA is high mainly for its \textit{inefficiency} in 
signal amplification, and \textit{non-linearity} in signal amplification 
outside the limited dynamic range. As a result, if the PA input signal is 
higher than its linear region, the output signal is distorted; 
whereas, if PA input signal falls below from a saturation point---where the 
output power is maximum---the PA drain efficiency drops significantly. It is 
energy efficient to operate the PA at its saturation point~\cite{PA_Survey}, 
but because of the dynamic range of input signals (i.e., higher-order modulations 
with high PAPR), it cannot operate at the saturation point and the PA must 
{back off} at {a} certain output power.

In earlier studies, a PA model with constant drain efficiency is assumed 
(e.g., see \cite{cui2005energy, wu2014energy}), which is far from reality. To 
capture the trade-off in energy efficiency and reliability, the link dynamics 
under realistic PA models must be studied. In the following, we compare the constant PA 
(CPA) with two non-linear PA models: a) commonly used 
traditional PA (TPA) and, b) envelop tracking PA (ETPA) that uses a linear 
PA along with a supply modulation circuitry in which the supply voltage 
tracks the input signal envelope.

\fakepar{Constant PA} The drain efficiency $\eta$ of CPA is assumed to be constant irrespective of 
the output power $P_t$  i.e., $\eta(P_t)=\eta_0$. 

\fakepar{Traditional PA}
Let $\eta(P_t)$ be the {drain} or power efficiency of PA at {the} output power $P_t$, then it 
is given by \cite{han2007power}
\begin{equation}
\eta(P_t) = \eta_{\max} \left(\frac{P_{t}}{P_{t,\max}}\right)^{\!\!1/2}, 
\label{eq:tpa_efficiency}
\end{equation}
where $P_{t, \max}$ is the maximum designed output power and $\eta_{\max}$ 
is the maximum PA efficiency, which is achieved when $P_t=P_{t,\max}$. 

At maximum output power e.g., $P_t = P_{0}$, a PA must be able to handle peak power, therefore $P_{0} \leq  P_{t,\max}/\xi$. It means that, for instance, if PAPR of a modulation scheme is \SI{8}{\decibel} then $P_{0}$ must be \SI{8}{\decibel} less than $P_{t,\max}$.  

\fakepar{Envelop tracking PA}
For ET-PA, the power efficiency $\eta(P_t)$ is modeled as \cite{hossain2011impactPA}
\begin{equation}
\eta(P_{t}) = \eta_{\max} \frac{P_{t} \left(1+c\right)}{P_{t} + c P_{t, \max}},
\label{eq:etpa_efficiency}
\end{equation}
where $c=0.0082$ is a PA-dependent constant.

\begin{figure}[!t]
\captionsetup[subfloat]{farskip=0pt,captionskip=-3pt} 
    \centering
        \includegraphics[width=1\linewidth]{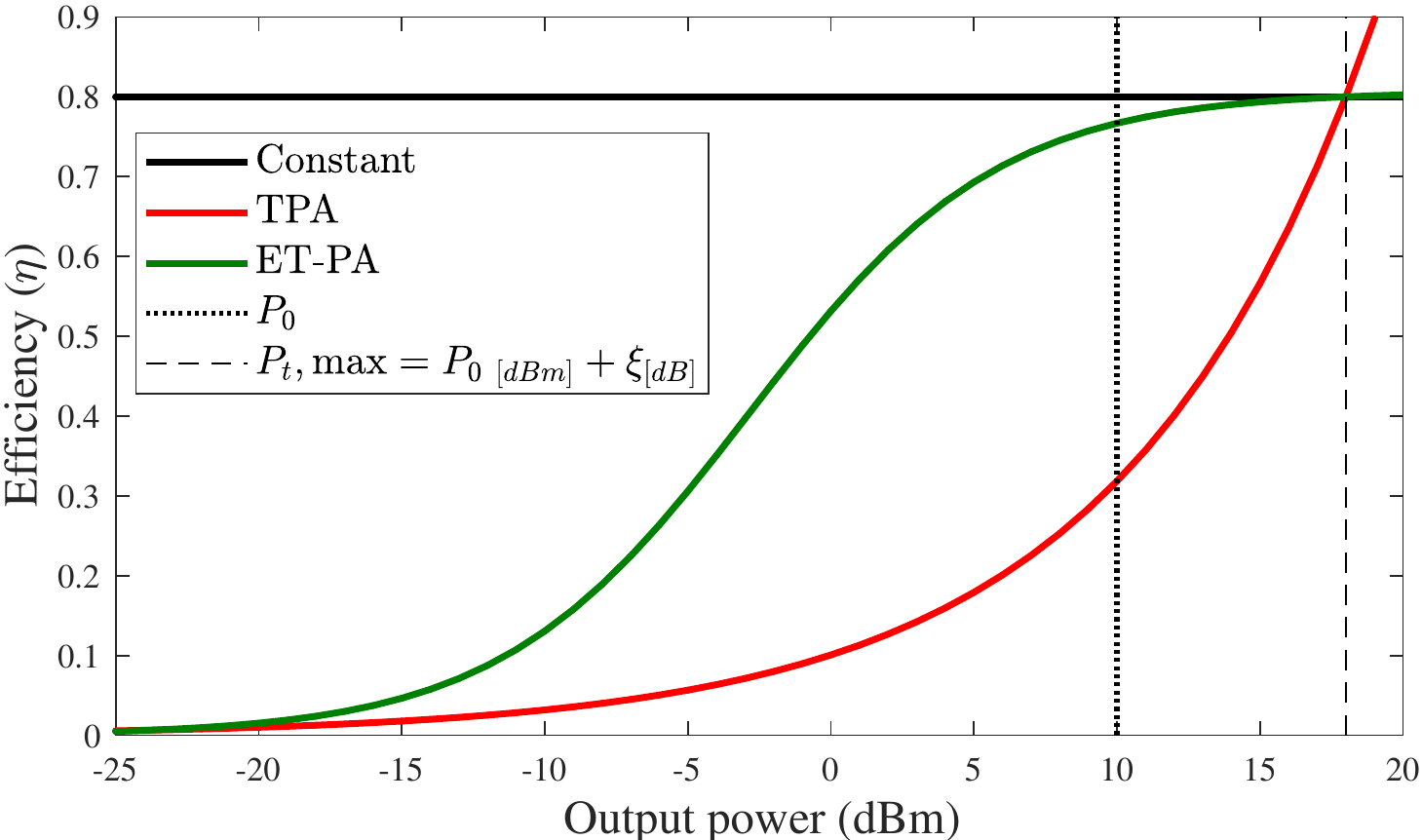}
  \caption{Efficiency curves of different PA models, assuming 
	$\eta_{\max}=\SI{80}{\percent}$ and $P_0 = \SI{10}{\decibelm}$.}
  \label{fig:EE_PA_Models} 
\end{figure}

Fig.~\ref{fig:EE_PA_Models} depicts the efficiency response of these PA 
models. It can be clearly observed how unrealistic CPA model is 
from real-life PA models, while ETPA model is expected to improve energy 
efficiency in comparison with TPA.   

Using these power efficiency relations, we derive energy consumption per 
information bit ($E_0$, \eqref{eq:E_0}). Table~\ref{tab:table_A} shows  
$E_0$ and its associated parameters for the consider PA models, where 
$R_b = W \log_2M$ is the PHY layer bit rate in bandwidth~$W$. 

\bgroup
\renewcommand*{\arraystretch}{2.7}
%  1 is the default, change whatever you need
\setlength{\tabcolsep}{3.5pt}
\begin{table}[!t]
%\small
\centering
 \caption{Expressions For Energy Per Information Bit for Studied Power Amplifier Models}
\label{tab:table_A}
\centering
\begin{tabular}{l|lll}
\noalign{\hrule height 1pt}
 & \textbf{CPA} & \textbf{TPA} & \textbf{ETPA}\\
\noalign{\hrule height 0.5pt}
$E_0$ & $\displaystyle\frac{n_p + n_h}{n_p} A\bar{\gamma} + B$ & $\displaystyle\frac{n_p + n_h}{n_p} A\frac{\bar{\gamma}}{\sqrt{\bar{\gamma}}} + B$ & Same as for \textbf{CPA} \\
$A$ 	& $\displaystyle\frac{\xi N_0 G_d}{\eta_0}$ & $\displaystyle\frac{\xi N_0 G_d\sqrt{P_{t, \max}}}{\eta_{\max}\sqrt{N_0 G_d R_b}}$ & $\displaystyle\frac{\xi N_0 G_d}{\eta_{\max} \left(c+1\right)}$\\
$B$ 	& $\displaystyle\frac{P_c}{R_b}$ & $\displaystyle\frac{P_c}{R_b}$ & $\displaystyle\frac{1}{R_b}\!\!\left(\!\frac{\xi c P_{t, \max}}{\eta_{\max}(c+1)}\! +\! P_c\!\!\right)$ \\[0.5em]
\noalign{\hrule height 1pt}
\end{tabular}
\end {table}
\egroup

%%%%%%%%%%%%%%%%\bgroup
%%%%%%%%%%%%%%%%%\def\arraystretch{2.7}
%%%%%%%%%%%%%%%%\renewcommand*{\arraystretch}{2.7}
%%%%%%%%%%%%%%%%%  1 is the default, change whatever you need
%%%%%%%%%%%%%%%%%\setlength{\tabcolsep}{3.5pt}
%%%%%%%%%%%%%%%%\begin{table*}[!t]
%%%%%%%%%%%%%%%%%\small
%%%%%%%%%%%%%%%%\centering
 %%%%%%%%%%%%%%%%\caption{Expressions For Energy Per Information Bit for Studied Power Amplifier Models}
%%%%%%%%%%%%%%%%\label{tab:table_A}
%%%%%%%%%%%%%%%%\centering
%%%%%%%%%%%%%%%%\begin{tabular}{l|lll}
%%%%%%%%%%%%%%%%\noalign{\hrule height 1pt}
 %%%%%%%%%%%%%%%%& \textbf{CPA} & \textbf{TPA} & \textbf{ETPA}\\
%%%%%%%%%%%%%%%%\noalign{\hrule height 0.5pt}
%%%%%%%%%%%%%%%%$E_0$ & $\dfrac{n_p + n_h}{n_p} A\bar{\gamma} + B$ & $\dfrac{n_p + n_h}{n_p} A\dfrac{\bar{\gamma}}{\sqrt{\bar{\gamma}}} + B$ & Same as for \textbf{CPA} \\
%%%%%%%%%%%%%%%%$A$ 	& $\dfrac{\xi N_0 G_d}{\eta_0}$ & $\dfrac{\xi N_0 G_d\sqrt{P_{t, max}}}{\eta_{\max}\sqrt{N_0 G_d R_b}}$ & $\dfrac{\xi N_0 G_d}{\eta_{\max} \left(c+1\right)}$\\
%%%%%%%%%%%%%%%%$B$ 	& $\dfrac{P_c}{R_b}$ & $\dfrac{P_c}{R_b}$ & $\dfrac{1}{R_b}\left(\dfrac{\xi c P_{t, \max}}{\eta_{\max}(c+1)} + P_c\right)$ \\[0.5em]
%%%%%%%%%%%%%%%%\noalign{\hrule height 1pt}
%%%%%%%%%%%%%%%%\end{tabular}
%%%%%%%%%%%%%%%%\end {table*}

\subsection{Packet Error Rate}
\label{subsec:PER}
To minimize energy consumption per information bit $E$ in \eqref{eq:E},
we require a generic packet error rate (PER) approximation that is accurate and also 
maintains an explicit connection with the parameters defining the 
system performance. An accurate approximation for uncoded schemes is proposed 
in \cite{Our2016}, which however cannot be utilized for optimizing the function $E$. Another, 
approximation is the upper bound on PER for (un)coded schemes in Rayleigh 
fading  
\begin{equation}
\bar{P_e}\left(\bar{\gamma}\right) \leq 1-\mathrm{exp}\left(-\frac{\omega_0}{
\bar{
\gamma}}\right), 
\label{eq:UB_Ray}
\end{equation}
where $\bar{\gamma}$ is the SNR and $\omega_0$
 is the waterfall threshold \cite{xi2011general}. The threshold is defined as 
an integral of the PER function in the AWGN channel, $f\left(\gamma\right)$
\begin{equation}
\omega_0 = \int_0^\infty {f\left(\gamma\right)d\gamma}. 
\label{eq:threshold_0}
\end{equation} 
For an $N$-bit uncoded packet with {a} bit error rate (BER) 
function $b_e\left(\gamma\right)$, $f\left(\gamma\right)$ is defined as
\begin{equation}
%\vspace{-2pt}
f(\gamma) = 1-\Big(1-b_e\left(\gamma\right)\Big)^{N}.
\label{eq:p_awgn1}
\end{equation}  

A log-domain linear approximation of $\omega_0$ is proposed for uncoded 
schemes in \cite{liu2012tput}, and for (un)coded schemes in 
\cite{wu2014energy}. However, the approximation in \cite{liu2012tput} is 
tight for large packets only while in \cite{wu2014energy} the approximation 
parameters for a given modulation scheme are calculated by simulations. 
%\st { A tight approximation to 
%the waterfall threshold and the PER in Nakagami-$m$ fading channels is one of 
%the contributions in this paper.}
On the other hand, the following new proposition shows that the waterfall threshold in 
\eqref{eq:threshold_0} is tightly approximated by the expected value of an 
asymptotic distribution of $f(\gamma)$. 

%In this paper, we propose simple approximations of $\omega_0$ and $\omega_m$ 
%that can accurately parametrize the PER upper bound for commonly-used uncoded 
%schemes. Uncoded schemes are attractive in particular for low-power radio 
%systems, and optimizing their energy-aware operation requires a PER 
%expression that can explicitly build a connection between the physical and 
%medium access layer parameters. To this end, we show that $\omega_0$ is 
%tightly approximated by the expected value whereas $\omega_m$ by the 
%\textit{m}th moment of the asymptotic distribution of the PER in the AWGN channel. To 
%the authors' best knowledge this is the first study that gives a closed-form 
%approximation of the threshold for Nakagami-$m$ fading channels. We also 
%reformulate the parameters of the asymptotic distribution that in 
%turn yield the tractable PER expressions in terms of elementary functions.   
  
%
%In what follows, we propose generic approximations to $\omega_0$
 %and $\omega_m$ for uncoded schemes with BER functions as
%\begin{equation} 	
%b_{e}(\gamma) = c_m\mathrm{exp}\left(-k_m\gamma\right) 
%\label{eq:f_se}
%%\vspace{-6pt} 
%\end{equation} 
%\begin{equation} 	
%b_{e}(\gamma) = c_mQ\left(\sqrt{k_m\gamma}\right) 
%\label{eq:f_sq}
%%\vspace{-4pt} 
%\end{equation} 
%where $c_m$ and $k_m$ are modulation-dependent constants. 
%Non-coherent FSK 
%and DPSK have the BER in the form of \eqref{eq:f_se} while M-ASK, M-PAM, 
%MSK, M-PSK and M-QAM have BER in the Gaussian $Q$-function form \eqref{eq:f_sq}\cite{xi2011general}.

\textit{Proposition 1}: For uncoded transmission of a packet with length $N$, 
with a BER function described by $c_m e^{-k_m\gamma}$ or 
$c_m Q(\sqrt{k_m\gamma})$ where $0 < c_m \leq 1$ and $k_m >0$, the 
threshold, $\omega_0$, in Rayleigh fading channel is approximated 
by the expected value of the Gumbel distribution for sample maximum  
\begin{equation}
\omega_0 \approx {\mathbb{E}[\gamma]} \triangleq a_N +  \gamma_e \, b_N,
\label{eq:proposedThreshold}
\end{equation}
where $a_N$ and $b_N$ are the parameters of the Gumbel distribution, and $\gamma_e$ is the Euler constant.

\textit{Proof}: See Appendix. \QEDA

In \cite{Our2016}, we have shown that the normalizing constants for BER 
functions of the form $c_m e^{-k_m\gamma}$ and $c_m Q(\sqrt{k_m\gamma})$ are 
\begin{equation}
a_N = \frac{\log (Nc_m)}{k_m},\,\,\,\,\,\,\,\, b_N = \frac{1}{
k_m},
\label{eq:AnBn_expo}
\end{equation}
where the optimal parameters for the BER function $c_m Q(\sqrt{k_m\gamma})$ 
are $c_m \Rightarrow 0.2114 c_m$ and $k_m \Rightarrow 0.5598 k_m$~\cite{MahmoodWCNC18}.

%One can find the \textit{m}th moment of the Gumbel distribution from its 
%moment generating function (MGF) defined as
%\begin{equation}
%M_{\gamma}\left(t\right) \triangleq \Gamma \Big(1-b_N \,t\Big)\,e^{{a_N \,t}}  
%\label{eq:MGF_Gumbel}
%\end{equation} 
%where $\Gamma\left(\cdot\right)$ is the standard gamma function. In Rayleigh 
%fading with $m=1$, from \eqref{eq:proposedThreshold} and 
%\eqref{eq:MGF_Gumbel}, $\omega_0$ equals the expected value of the Gumbel 
%distribution, i.e.
%\begin{equation}
%\omega_0 \approx \mathbb{E}[\gamma] = a_N +  \gamma_e \, b_N
%\label{eq:solution_w0}
%\end{equation}

Inserting \eqref{eq:proposedThreshold} and \eqref{eq:AnBn_expo} in 
\eqref{eq:UB_Ray} leads to a simple PER approximation which maintains 
explicit connection with the modulation specific parameters ($c_m$, $k_m$), packet size ($N$) and SNR ($\bar{\gamma}$) 
\begin{equation}
\bar{P}_{e}(\bar{\gamma}) \approx 1-\big(N\big)^{-\frac{1}{
k_m\bar{\gamma}}} \mathrm{exp}\left(-\frac{\log c_m + \gamma_e}{k_m\bar{
\gamma}}\right).
\label{eq:ubPER1}
\end{equation}
%
 %which is 
%accurate than \cite{wu2014energy}\cite{liu2012tput} and also maintains 
%explicit connection with the modulation order, unlike \cite{wu2014energy}

\begin{figure}[!t]
\captionsetup[subfloat]{farskip=0pt,captionskip=-3pt} 
    \centering
        \includegraphics[width=1\linewidth]{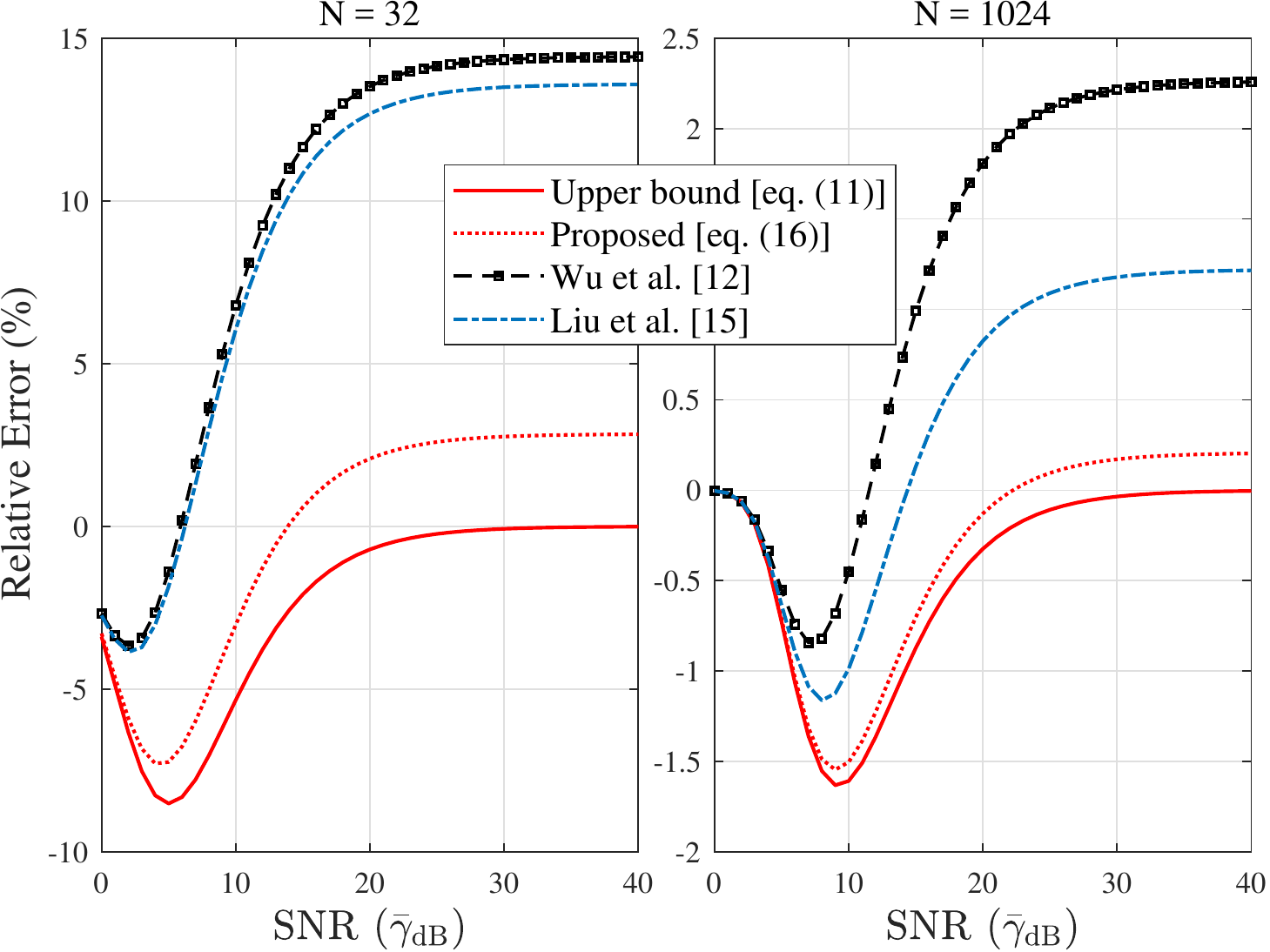}
  \caption{Relative error (RE) in average PER for 16 QAM modulation in Rayleigh fading.}
  \label{fig:16QAM} 
\end{figure}

The PER approximation in \eqref{eq:ubPER1} tightly matches to the upper bound, as 
compared to existing solutions for any modulation scheme with the BER function 
in the form of exponential and the Gaussian-$Q$ function. In order to 
show the accuracy of proposed approximation, using as an example of uncoded 16 QAM, 
Fig.~\ref{fig:16QAM} compares it with the numerical evaluation of 
\eqref{eq:UB_Ray}, and also with the prior approximations in terms of relative error (RE) with respect 
to real PER. {The real PER is found by the numerical integration of \eqref{eq:p_awgn1} over Rayleigh fading distribution.} It can be observed that the RE of the proposed approximation is close to the upper 
bound \eqref{eq:UB_Ray} for small to large packet lengths. In comparison, the 
RE of approximations in \cite{wu2014energy}\cite{liu2012tput} is small at low SNR, however it 
increases rapidly especially for small packet lengths.

\fakepar{PER for Coded Schemes}
We can easily use the approximation \eqref{eq:ubPER1} for PER 
evaluation of coded schemes by equating $k_M = 1/k_m$, and 
$b_M = \left(\log c_m + \gamma_e\right)/k_m$, where  $k_M$ and $b_M$ are 
calculated by simulations for various coded schemes in 
\cite[Table~I]{wu2014energy}. Therefore, one can perform the 
energy-reliability analysis presented in this paper for coded schemes 
{as well}. However, as the low-cost and low-power IoT devices rely on 
retransmissions instead of error-correction to reduce the system complexity
\cite{Tania_Alouini}, we adhere to uncoded schemes to be 
compatible with low-cost IoT devices.

%If \eqref{eq:PER_req} is maintained for each transmission, the $\tau_r^{\max}$
%at the data link layer will satisfy the delay and reliability targets.

%%%%%%%%%%%%%%%%%%%%%%%%%%%%%%%%%%%%%%%%%%%%%%%%%%%%%%%%%%%%%%%%%%%%%%%%%%%
%%%%%%%%%%%%%%%%%%%%%%%%%%%%%%%%%%%%%%%%%%%%%%%%%%%%%%%%%%%%%%%%%%%%%%%%%%%
%%%%%%%%%%%%%%%%%%%%%%%%%%%%%%%%%%%%%%%%%%%%%%%%%%%%%%%%%%%%%%%%%%%%%%%%%%%
%%%%% Link Optimization with Minimum Energy Consumption
%%%%%%%%%%%%%%%%%%%%%%%%%%%%%%%%%%%%%%%%%%%%%%%%%%%%%%%%%%%%%%%%%%%%%%%%%%%
%%%%%%%%%%%%%%%%%%%%%%%%%%%%%%%%%%%%%%%%%%%%%%%%%%%%%%%%%%%%%%%%%%%%%%%%%%%
%%%%%%%%%%%%%%%%%%%%%%%%%%%%%%%%%%%%%%%%%%%%%%%%%%%%%%%%%%%%%%%%%%%%%%%%%%%

\section{Link Optimization with Minimum Energy Consumption}
\label{sec:EE}
With two main components of our objective function $E$ in place, now we 
find energy optimal yet QoS-aware (i.e., maintaining the PER constraint in 
\eqref{eq:PER_req}) PHY and MAC parameters. 

%In next section, we consider minimizing energy consumption per information 
%bit in \eqref{eq:E} while maintaining the PER constraint in 
%\eqref{eq:PER_req}.

\subsection{Optimal Average SNR / Transmit Power}
To find an energy-optimal SNR, we fix the packet payload $n_p$. It also 
represents a case where the sensors have to send fixed-size reports. 
Moreover, the optimal SNR must satisfy the reliability constraints 
set via PER and the constraints on the maximum transmit power, which could be 
either due to hardware limitations or frequency band regulations. Therefore, 
the optimization of objective function \eqref{eq:E} with respect to SNR $\bar{
\gamma}$ can be written as
%With $n_p$ fixed, finding the optimal average SNR represents a case where the 
%sensors have to send a fixed size reports.  
%Under the constraints on required PER and the transmit 
%power, the minimization of energy in \eqref{eq:E} can be written as
\begin{equation}
\begin{aligned}
& \underset{\bar{\gamma}}{\text{minimize}}
& & E(\bar{\gamma}) \\[-0.3em]
& \text{subject to}
& & \bar{\gamma}_{\min}\leq\bar{\gamma} \leq \bar{\gamma}_{\max},
\end{aligned}
\label{eq:Const_snr}
\end{equation}
where $\bar{\gamma}_{\min}$ is the minimum average SNR requirement under PER 
bound in \eqref{eq:PER_req}. From \eqref{eq:ubPER1}, $\bar{\gamma}_{\min}$ is
\begin{equation}
\bar{\gamma}_{\min} = - \frac{\gamma_e + \log\Big(c_m\left(n_h+n_p\right)
\Big)}{k_m \log\left(1-\mathcal{P}_{\mathrm{req}}\right)}.
\label{eq:snr_min}
\end{equation}
While $\bar{\gamma}_{\max}$ is the maximum achievable SNR at transmit power limit $P_0$. Since the transmit power $P_t$ cannot exceed $P_0$, i.e., $P_{\mathrm{t}} \leq P_0$, it translates to 
$\bar{\gamma} \leq \bar{\gamma}_{\max}$ with $\bar{\gamma}_{\max}$ from \eqref{eq:E_t} as
\begin{equation}
\bar{\gamma}_{\max} = \frac{P_0}{WN_0G_d}. %\log_2 M 
\label{eq:snr_max}
\end{equation}

From \eqref{eq:snr_min} and \eqref{eq:snr_max}, the required SNR, denoted as 
$\bar{\gamma}^{*}_{\mathrm{req}}$, relates to an {unconstrained optimal SNR $\bar{\gamma}^{*}$} as 
\begin{align}
\bar{\gamma}^{*}_{\mathrm{req}}=
 \begin{dcases}
\bar{\gamma}_{\min}, & \bar{\gamma}^{*} < \bar{\gamma}_{\min}
\\
\bar{\gamma}_{\max}, & \bar{\gamma}^{*} > \bar{\gamma}_{\max}\\
\,\, \bar{\gamma}^{*}, &  \mathrm{otherwise}\\
 \end{dcases}
\label{eq:H_n_x}
\end{align} 
which holds only if the condition $\bar{\gamma}_{\min} < \bar{\gamma}_{\max}$ 
is satisfied. Otherwise, the reliability target cannot be satisfied for a 
given modulation scheme because of transmit power constraint. 

\begin{figure}[t]
	\centering
		\includegraphics[width=0.95\linewidth]{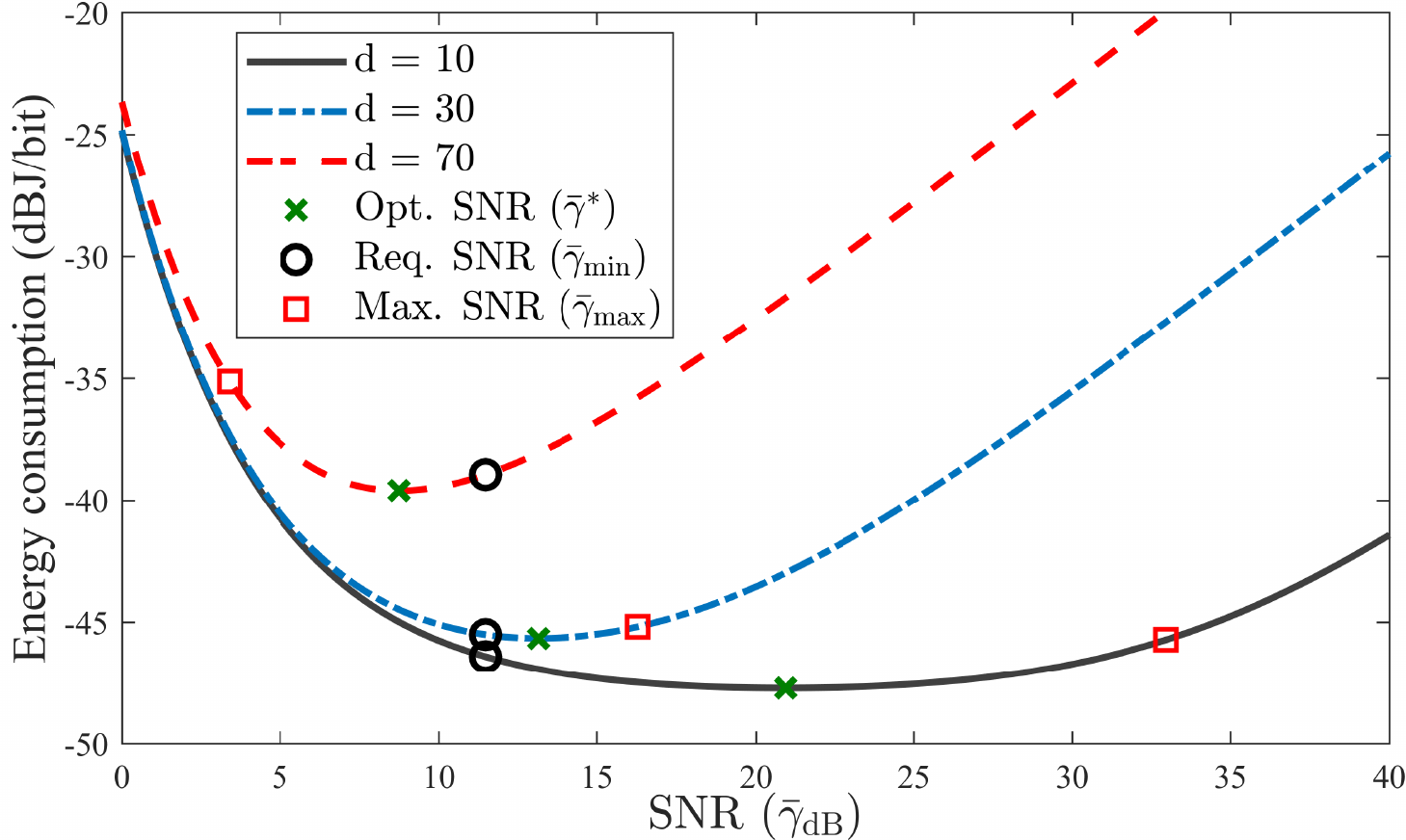}
		%\vspace{-18pt}
	\caption{Comparison of optimal, required and maximum achievable SNRs at different distances ($d$) for 4-QAM under reliability constraints of $
\mathcal{P}=\num{0.001}$, $\tau_r^{\max}=\num{3}$, $n_p = \SI{976}{\bit}$ and maximum transmit power 
$P_0=\SI{10}{\milli\watt}$.}
	\label{fig:opt_snr}
	%\vspace{-14pt}
\end{figure}

The conditions in \eqref{eq:H_n_x} can be visualized with 
Fig.~\ref{fig:opt_snr}, {which is} obtained using the parameters in 
Sec.~\ref{sec:Results}~Table.~\ref{tab:Tab2} under constant PA. Fig.~\ref{fig:opt_snr} shows an 
example case of minimum required SNR $\bar{ \gamma}_{\min}$ to satisfy 
certain QoS at various distances. It also depicts the unconstrained optimal 
SNR $\bar{\gamma}^*$ and maximum achievable SNR 
$\bar{\gamma}_{\max}$. At $d=\SI{10}{\meter}$, $\bar{\gamma}_{\min}$ is less 
than $\bar{\gamma}^*$, therefore $\bar{\gamma}^* $ is energy optimal and is 
preferred over $\bar{\gamma}_{\min}$. While at $d=\SI{30}{\meter}$, 
$\bar{\gamma}^*$ cannot satisfy the target and $\bar{\gamma}_{\min}$, though 
not energy optimal, is selected. At $d=\SI{70}{\meter}$, the reliability 
target is not satisfied as $\bar{\gamma}_{\min}> \bar{\gamma}_{\max}$.

We can find $\bar{\gamma}^{*}$, which is finally conditioned by 
\eqref{eq:H_n_x}, based on the following unconstrained optimization problem
\begin{equation}
\begin{aligned}
& \underset{\bar{\gamma}}{\text{minimize}}
& & E(\bar{\gamma}) \\[-0.3em]
& \text{subject to}
& & \bar{\gamma} \in \left[0, \infty\right].
\end{aligned}
\label{eq:unConst_snr}
\end{equation}
Note that $E(\bar{\gamma})$ is a product of two functions: the 
number of retransmissions $\bar{\tau}(\gamma)$ with $\bar{\tau}^{'}(\gamma)\leq 0$, and 
the average energy per transmission attempt $E_0(\gamma)$ such that 
$E_0^{'}(\gamma)\geq0$ where $\acute{x}$ denotes the first derivative. Given that both $\bar{\tau}(\gamma)$ and $E_0(\gamma)$ are convex, the function $E(\bar{\gamma})$ is also convex \cite[Lemma~1]{wu2014energy}. 

\fakepar{CPA/ETPA model} For CPA/ETPA model with $E_0$ in 
Table~\ref{tab:table_A}, $\bar{\gamma}^{*}$ can be obtained by solving 
$\frac{\partial E} {\partial \bar{\gamma}} = 0$ in \eqref{eq:unConst_snr}, which yields a quadratic equation 
with a positive root as 
\begin{equation}
\bar{\gamma}^{*} = \frac{\omega_0}{2} + \sqrt{\omega_0 \left(\frac{\omega_0}{4
} + \frac{B}{A}\frac{n_p}{n_h+n_p}\right)}.
\label{eq:opt_snr}
\end{equation}

\fakepar{TPA model} For $E_0$ with TPA model in Table~\ref{tab:table_A}, although $E$ is convex, the optimal $\bar{\gamma}^*$ is not found explicitly. However, it can be determined numerically from the following non-quadratic equation
\begin{equation}
A k_m \left(\bar{\gamma}\right)^{3/2}  - 2A\omega_0 (\bar{\gamma})^{1/2}- {2\omega_0B}\left(\frac{n_p}{{n_p+n_h}}\right)=0,
\label{eq:opt_snr_tpa}
\end{equation}
which has a real positive root if the condition $\!27 B^2  k_m \left({n_p}/{(n_p\!+\!n_h)}\right)^{2} > 8 A^2\omega_0$ is satisfied
\begin{equation}
\bar{\gamma}^{*} = \frac{1}{3} \left( \frac{4 \omega_0}{k_m} \,\, + \,\, \frac{4 \omega_0^2}{\mathcal{A}^{\frac{1}{3}}} \,\, + \,\, \frac{\mathcal{A}^{\frac{1}{3}}}{k_m^2}\right),
\end{equation}
where 
\begin{align}
\mathcal{A}  = & \frac{1}{A^2} \bigg[54B^2 k_m^4 \omega_0^2 \Big(\frac{n_p}{n_p\!+\!n_h}\Big)^{\!\!2} - 8A^2k_m^3 \omega_0^3 + 6\sqrt{3} \,\,\times \nonumber \\ 
							& \!\!\!\left(\!B^2 k_m^7 \omega_0^4 \Big(\frac{n_p}{n_p\!+\!n_h}\Big)^{\!\!2} \Big(\!27 B^2  k_m \Big(\frac{n_p}{n_p\!+\!n_h}\Big)^{\!\!2} \! - \! 8 A^2\omega_0\!\Big)\!\right)^{\!\frac{1}{2}}\!\bigg]. \nonumber
\end{align}

\subsection{Optimal Packet Payload} 
Now we find the optimal $n_p$---the payload that minimizes the 
objective function \eqref{eq:E}---by keeping SNR constant. The 
upper limit on the maximum payload size ${n_{p,\max}}$ is determined by 
required
$\bar{\gamma}_{\min}$ to satisfy a PER target. Therefore, our objective 
function is 
\begin{equation}
\begin{aligned}
& \underset{\bar{\gamma}}{\text{minimize}}
& & E(n_p) \\[-0.3em]
& \text{subject to}
& & 0 \leq n_p \leq n_{p,{\max}},
\end{aligned}
\label{eq:Const_np}
\end{equation}
where $n_{p,{\max}}$, from \eqref{eq:PER_req}, is
\begin{equation}
{n_{p,\max}} = - n_h + \frac{1}{c_m}10^{\displaystyle{-\big(\gamma_e + \bar{
\gamma}_{\min} k_m \log\left(1-\mathcal{P}_{\textrm{req}}\right)\big)}},
\label{eq:max_L}
\end{equation}
with $\bar{\gamma}_{\min}$ defined in \eqref{eq:snr_min}. After 
reliability condition known with respect to maximum payload size, $n_p^{*}$ obtained 
via  unconstrained optimization of $E(n_p)$ can easily be conditioned i.e., if 
$n_p^{*} > n_{p,{\max}}$ then $n_p= n_{p,{\max}}$ and otherwise $n_p = n_p^{*}$. 

The function $E(n_p)$ is convex in $n_p$ and the unconstrained optimal $n_p^{*}$ 
for the considered PA models is given below.

%the solution of $\frac{\partial E}{\partial n_p} = 0$ as 

%The upper limit on the payload size ${n_{p,\max}}$ is set by the minimum SNR requirement $\bar{\gamma}_{\min}$ to satisify PER target. It is given by from \eqref{eq:PER_req}

\fakepar{CPA/ETPA model} In case of ETPA, the optimal $n_p^*$ is determined as
\begin{equation}
n_p^{*} = \!\!\frac{n_h \bar{\gamma}\left(\left(k_m\! -1\right)\bar{\gamma}\! + \!\sqrt{
k_m^2\bar{\gamma}^2\! + \! 2 k_m \bar{\gamma}\! + \!\frac{4Bk_m}{A}\! + \!1}\right)}{2\Big(
\bar{\gamma} + \frac{B}{A}\Big)}.
\label{eq:opt_L_etpa}
\end{equation}

\fakepar{TPA model} For TPA, $n_p^*$ is
\begin{equation}
n_p^{*} = \frac{n_h\mathcal{K} + \sqrt{4Ak_mn_h^2\bar{\gamma}\left(A\bar{\gamma}-B\sqrt{\bar{\gamma}}\right)\left(\bar{\gamma}-\frac{B^2}{A^2}\right)- n_p^2\mathcal{K}^2}}{2A\left(\bar{\gamma}-\frac{B^2}{A^2}\right)},
\label{eq:opt_L_tpa}
\end{equation}
where $\mathcal{K} = Ak_m\bar{\gamma}^2 - B k_m \bar{\gamma}^{3/2} -A\bar{\gamma}+B\sqrt{\bar{\gamma}}$.

%-L0 + (10^(-(C + snr_min*Km*k2*log(1-PER_req))))/(Cm*k1)

\subsection{Joint Energy Optimal Parameters---$\bar{\gamma}, n_p, M, \tau_{r}^{\max}$} 

As the IoT devices will be used in diverse monitoring and control 
applications, it might be important in many to find the optimal SNR, payload 
size, modulation order and the number of retransmissions for energy efficient 
communication. For example, after 
deployment in harsh and inaccessible areas, the devices can optimize these 
parameters at the start of their operation and then continue with the 
optimal link setting. The problem to jointly optimization these parameters 
can be formulated as 
\begin{equation}
\begin{aligned}
& \underset{\bar{\gamma}, n_p, M, \tau_{r}^{\max}}{\text{minimize}}
& & E\Big(\bar{\gamma}, n_p, M, \tau_{r}^{\max}\Big),
%& \text{subject to}
%& & \bar{\gamma}_{\min}\leq\bar{\gamma} \leq \bar{\gamma}_{\max}\\
%&&& M \in \left[\mathrm{FSK, MPSK, MQAM}\right]\\
%&&& \tau_{r}^{\max} = i, \; i = 1, \ldots, m.
\label{eq:JointOptimal}
\end{aligned}
\end{equation}
where $M \in \{\mathrm{FSK, MPSK, MQAM}\}$ and $\tau_{r}^{\max} = i, \; i \geq 1$.
%In order to solve this optimization problem, 
%In general solving this optimization problem is {need to 
%motivate the numerical and iterative process}, 
%
%
%however, any of these IoT 
%devices will support only few  types of $M$, and a small value of 
%$\tau_{r}^{\max}$ is feasible for minimum energy operation 
%\cite{wu2014energy}. 
Note that the IoT devices will support only a few modulation {schemes} $M$. In addition, a small value 
of maximum retransmissions $\tau_{r}^{\max}$ is feasible for operation with minimum energy consumption
\cite{wu2014energy}. As a result, an exhaustive search over the combination 
of $M$ and $\tau_{r}^{\max}$ will not be computationally demanding. 

On the other hand, for each combination of $M$ and $\tau_{r}^{\max}$, the joint optimum of $\bar{\gamma}$ 
and $n_p$ can be found from \eqref{eq:opt_snr} and \eqref{eq:opt_L_etpa} either by 
solving a system of two non-linear equations or by iteratively invoking these 
equations. In either case, we must ensure that the reliability conditions 
in \eqref{eq:H_n_x} and \eqref{eq:max_L} are satisfied. However, the former 
approach requires numerical evaluation that might not be computationally feasible 
for hardware-constrained IoT devices. Whereas, by iteratively invoking 
\eqref{eq:opt_snr} and \eqref{eq:opt_L_etpa}, $\bar{ \gamma}$ and $n_p$ can 
efficiently converge to joint energy optimum values while satisfying the 
reliability conditions. It is straightforward to develop the proof of 
convergence of the iterative approach~\cite[Corollary 3]{wu2014energy} under the 
probabilistic QoS model. We observed that by initializing $n_p$ and $\bar{\gamma}$ 
to any random values, this approach converges to their optimum values within a few 
iterations.

{A pseudocode of the proposed joint optimization of 
\eqref{eq:JointOptimal} is given in Algorithm~1. At a given distance, the 
algorithm iterates over all combinations of $M$  and $\tau_{r}^{\max}$ while 
for each combination, $\bar{ \gamma}$ and $n_p$ are 
evaluated iteratively. The convergence of $\bar{ \gamma}$ and $n_p$ is monitored 
with the residual SNR $\Delta > \delta$, where $\delta$ is the precision 
tolerance.}   

\setlength{\textfloatsep}{4pt}
% Remove \textfloatsep
\begin{algorithm}[!t]
\small
 \caption{Joint Energy-Reliability Aware Optimization of a Link's Parameters}
 \begin{algorithmic}[1]
 \renewcommand{\algorithmicrequire}{\textbf{Input:}}
 \renewcommand{\algorithmicensure}{\textbf{Output:}}
 \REQUIRE $\mathcal{P}_{\mathrm{req}}, \tau_r^{\max}, \delta$
 \ENSURE  $\bar{\gamma}^{*}, n_p^*, \tau_r^*, M^*$
		\FOR {$M \in \left[\mathrm{FSK, MPSK, MQAM}\right]$}
			\FOR {$i = 1$ to $\tau_r^{\max}$}
				\STATE $n_p \gets 0$	
					\WHILE {$\Delta > \delta$}
						%\STATE $\bar{\gamma} \gets \bar{\gamma}^{*}, n_p \gets n_p^*$
						\STATE  $\bar{\gamma} \gets$ Evaluate \eqref{eq:opt_snr} or \eqref{eq:opt_snr_tpa}, $\bar{\gamma}_{\min} \gets$ Evaluate 
						\eqref{eq:snr_min}, \\ $\bar{\gamma}_{\max} \gets$ Evaluate \eqref{eq:snr_max}, $n_{p,{\max}} \gets$ Evaluate \eqref{eq:max_L}
						\IF {($\bar{\gamma}_{\min}> \bar{\gamma}_{\max}$)}
					       \STATE $\mathrm{break}$;
						\ELSE
					     	\STATE $ \bar{\gamma}_{\mathrm{req}} \gets $ Evaluate \eqref{eq:H_n_x} 
					 	\ENDIF
								\STATE  $n_p \gets$ Evaluate \eqref{eq:opt_L_etpa} or \eqref{eq:opt_L_tpa} with $\bar{\gamma}$ =  $\bar{\gamma}_{\mathrm{req}}$
            \IF {$(n_p>n_{p,{\max}})$}
						    \STATE $n_p \gets   n_{p,{\max}}$
					  \ENDIF
								\STATE 	$E  \gets$ Evaluate \eqref{eq:E_trunc}, Print $E,{\gamma}, n_p,\tau_r, M$
								\\  $\Delta  \gets |\bar{\gamma}_{\mathrm{req}} - \bar{\gamma}^{'}|, \,\, \bar{\gamma}^{'} = \bar{\gamma}_{\mathrm{req}}$
					\ENDWHILE
			\ENDFOR		
		\ENDFOR
 \RETURN $\bar{\gamma}, n_p, \tau_r, M$ giving minimum $E$
 \end{algorithmic}
%\vspace{-3pt} 
 \end{algorithm}

%%%%%%%%%%%%%%%%%%%%%%%%%%%%%%%%%%%%%%%%%%%%%%%%%%%%%%%%%%%%%%%%%%%%%%%%%%%
%%%%%%%%%%%%%%%%%%%%%%%%%%%%%%%%%%%%%%%%%%%%%%%%%%%%%%%%%%%%%%%%%%%%%%%%%%%
%%%%%%%%%%%%%%%%%%%%%%%%%%%%%%%%%%%%%%%%%%%%%%%%%%%%%%%%%%%%%%%%%%%%%%%%%%%
%%%%% Numerical Results
%%%%%%%%%%%%%%%%%%%%%%%%%%%%%%%%%%%%%%%%%%%%%%%%%%%%%%%%%%%%%%%%%%%%%%%%%%%
%%%%%%%%%%%%%%%%%%%%%%%%%%%%%%%%%%%%%%%%%%%%%%%%%%%%%%%%%%%%%%%%%%%%%%%%%%%
%%%%%%%%%%%%%%%%%%%%%%%%%%%%%%%%%%%%%%%%%%%%%%%%%%%%%%%%%%%%%%%%%%%%%%%%%%%
\section{Numerical Results}
\label{sec:Results}
%The simulation parameters are taken from \cite{cui2005energy}: $N_0/2 = \SI{-
%174}{\decibelm\per\hertz}$, $\kappa = \num{3.5}$, $G_1=\SI{30}{\decibel}$, $M_{
%\ell}=\SI{40}{\decibel}$, $W=\SI{10}{\kilo\hertz}$, 
%$P_c^{\textrm{\{MQAM, MPSK\}}}=\SI{310}{\milli\watt}$, $P_c^{\mathrm{FSK}}=\SI{
%265}{\mW}$, $\eta = \SI{35}{\percent}$.

 {In this section, we present the numerical results of the proposed link optimization approach. The 
parameters used for the numerical analysis are given in Table~\ref{tab:Tab2}}, where the PER 
target of $\mathcal{P} = \num{0.001}$ translates to \SI{99.9}{\percent} 
reliability.
%, selected according to
%\cite{cui2005energy}. 
%The other parameters are: $n_h = \SI{48}{\bit}$, $
%\varepsilon = \num{0.001}$ (i.e., \SI{99.9}{\percent} reliability). 

\bgroup
\def\arraystretch{1.0}
%  1 is the default, change whatever you need
%\setlength{\tabcolsep}{pt}
\begin{table}[!ht]
%\medium
\centering
 \caption{{Parameters for Numerical Analysis}}
\label{tab:Tab2}
\centering
\begin{tabular}{lll}
\noalign{\hrule height 1pt}
\textbf{Parameter}  & \textbf{Symbol} &\textbf{Value} \\
\noalign{\hrule height 0.7pt}
Max. transmit power        & $P_0$      & \SI{10}{\milli\watt} \\
Noise power density        & $N_0/2$    & \SI{-174}{\decibelm\per\hertz} \\
Path loss exponent         & $\kappa$   & \num{3.5} \\
Path gain at unit distance & $G_1$ 	    & \SI{30}{\decibel} \\
Link margin                & $M_{\ell}$ & \SI{40}{\decibel} \\
Bandwidth                  & $W$ 	      & \SI{10}{\kilo\hertz} \\
Circuit power - MQAM, MFSK & $P_c$      & \SI{310}{\milli\watt}, \SI{265}{\milli\watt} \\
%Circuit power - MFSK       & $-$       & \SI{265}{\milli\watt} \\
Max. PA efficiency         & $\eta_0, {\eta_{\max}}$   & \SI{80}{\percent} \\
Packet overhead            & $n_h$           & \SI{48}{\bit} \\
Target PER                 & $\mathcal{P}$   & \num{0.001}\\
Max. retransmission        & $\tau_r^{\max}$ & \num{3}\\
{Precision tolerance}   		 & {$\delta$}        & {$10^{-6}$} \\
\noalign{\hrule height 1pt}
\end{tabular}
\end {table}
\egroup
  
%\begin{figure}[t]
	%\centering
		%\includegraphics[width=1\linewidth]{Figures/OBS_L.eps}
		%%\vspace{-18pt}
	%\caption{}
	%\label{fig:obs_L}
	%%\vspace{-14pt}
%\end{figure}
%\begin{figure}[t]
	%\centering
		%\includegraphics[width=1\linewidth]{Figures/OBS_SNR.eps}
	%%\vspace{-18pt}
	%\caption{}
	%\label{fig:obs_snr}
%\end{figure}
%\fakepar{Energy Optimal SNR and Payload}
%
%\begin{figure*}[!ht] 
    %\centering
  %\subfloat[fig:obsL]{%
       %\includegraphics[width=0.49\linewidth]{Figures/OBS_L.eps}\label{fig:1a}}\hfill
  %\subfloat[fig:obssnr]{%
        %\includegraphics[width=0.49\linewidth]{Figures/OBS_SNR.eps}\label{fig:1b}}	
  %\caption{Coverage probability under studied outage conditions at $\eta = 3$ and two different cell sizes.}
  %\label{fig:obs} 
%\end{figure*}
%
%
%
%\begin{itemize}
	%\item If the SNR is kept the same at any distance, the optimal packet length
%increases slightly with the distance. AND $E_b$ also increases. WELL depends on the SNR
%value. If SNR is high and kept the same at any distance, then the packet 
%length increases significantly with the increase in distance.
 %\item If $L$ is kept the same at any distance, the optimal SNR decreases
%significantly. And $E_b$ also increases. 
%\item As with the increase in distance, the optimal SNR decreases, therefore,
%the optimal $L$ remains the same.
%\item The behavior remains the same, although energy consumption per bit is
%different, for the other power amplifier models.
%\end{itemize}

\begin{figure}[t]
	\centering
		\includegraphics[width=0.95\linewidth]{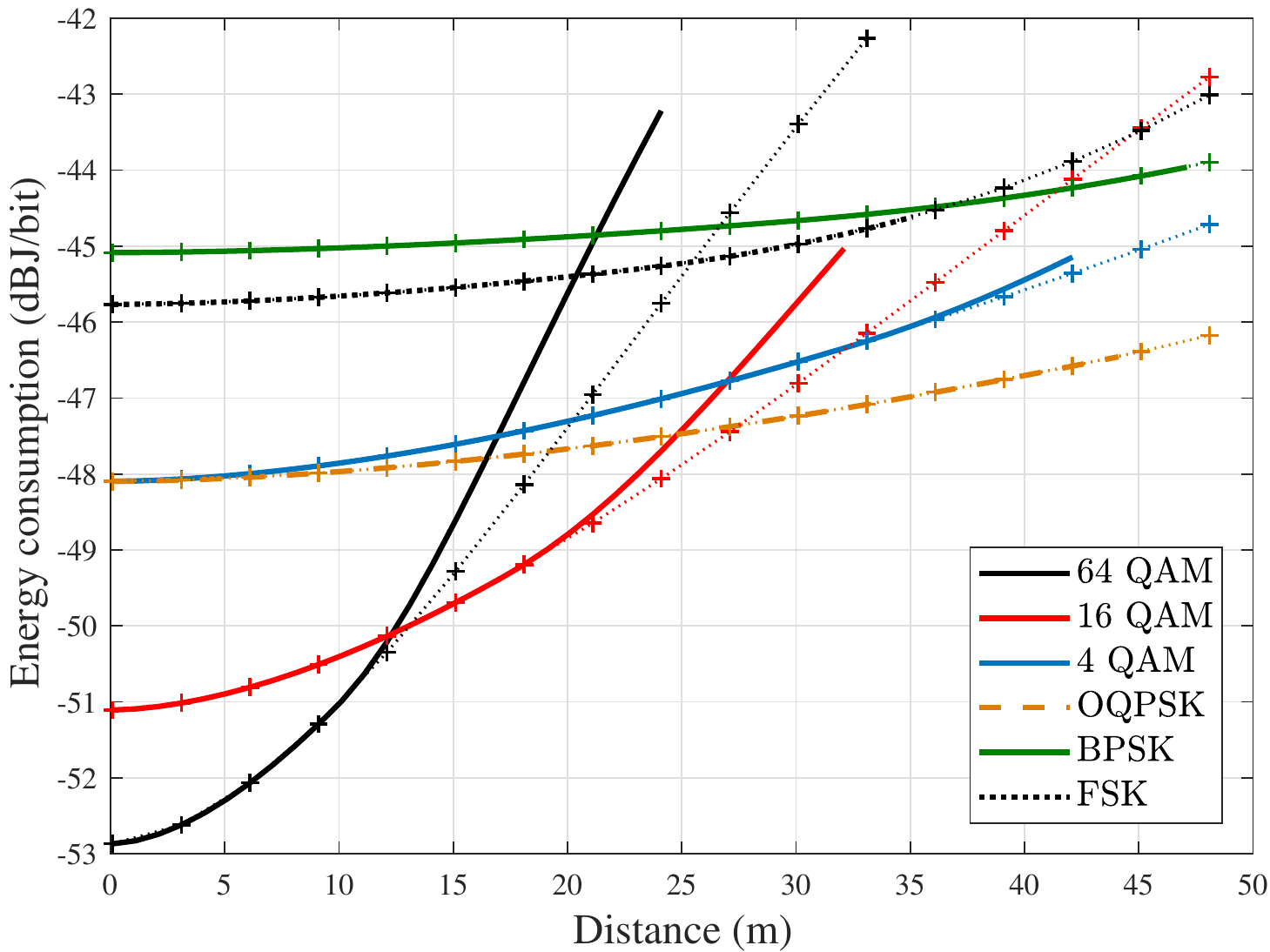}
%\vspace{-18pt}
	\caption{Energy consumption of various modulation schemes under CPA model 
with optimal SNR ($\bar{\gamma}^{*}$) and payload ($n_p^{*}$) at each 
distance. Reliability parameters: $\mathcal{P}=\num{
0.001}$, $\tau_r^{\max}=\num{3}$. 
%, Other parameters: $P_0=\SI{10}{\milli\watt}$, $n_h=\SI{48}{\bit}$, $\eta_{\max}=\SI{80}{\percent}$. 
The dotted-marker curves ($\cdot\cdot$+$\cdot\cdot$) show the energy consumption without any reliability constraints.}
	\label{fig:snr_d}
\end{figure}

When operated at {the} optimal SNR and payload size, the 
energy consumption (EC) of 
the selected modulation schemes with respect to distance with (solid lines) 
and without (dotted-marker lines) reliability constraints is 
shown in Fig.~\ref{fig:snr_d}. These results are obtained based on constant 
PA. It is observed that there is an energy-optimal modulation scheme at each 
distance that also satisfies the reliability target: higher-order modulations 
at lower distance and lower-order at higher distance. However, for a given 
transmit power limit, the distance until which the reliability target is 
satisfied decreases as the reliability requirements get tight. Also, how 
energy efficiency takes a toll under reliability constraints, compared to a 
link with unlimited allowed transmissions, can also be noticed especially 
for higher-order modulations. Note that the EC gap increases with 
reliability constraints becoming stringent. This is because, to meet the 
reliability target, the parameters other than the energy-optimal ones
 are selected. We observed 
the similar EC trend for other PA models however with 
exceptions that need {in-depth} analysis.
%In Fig.~\ref{fig:algo}, one can grasp the big picture of how the parameters $
%\bar{\gamma}, n_p, M$ and $\tau_{r}^{\max}$ vary with distance. At very short 
%distance high $M$ and lower $n_p$ are energy efficient. The reason behind 
%lower $n_p$ can be explained with the smaller value of $\tau_r^{\max}$. As 
%distance increases optimal $M$ becomes smaller. The payload size $n_p$ keeps 
%increasing at around $7-8$m and $13-19$m region with the increase in distance 
%keeping the $\bar{\gamma}$ almost constant, i.e., increasing transmit power 
%with increasing packet size is optimal until next smaller $M$ becomes energy 
%optimal. At long distances lower $M$ and $n_p$ are energy optimal.

\begin{figure*}[!ht] 
    \centering
  \subfloat[Energy consumption ($E$)]{%
       \includegraphics[width=0.50\linewidth]{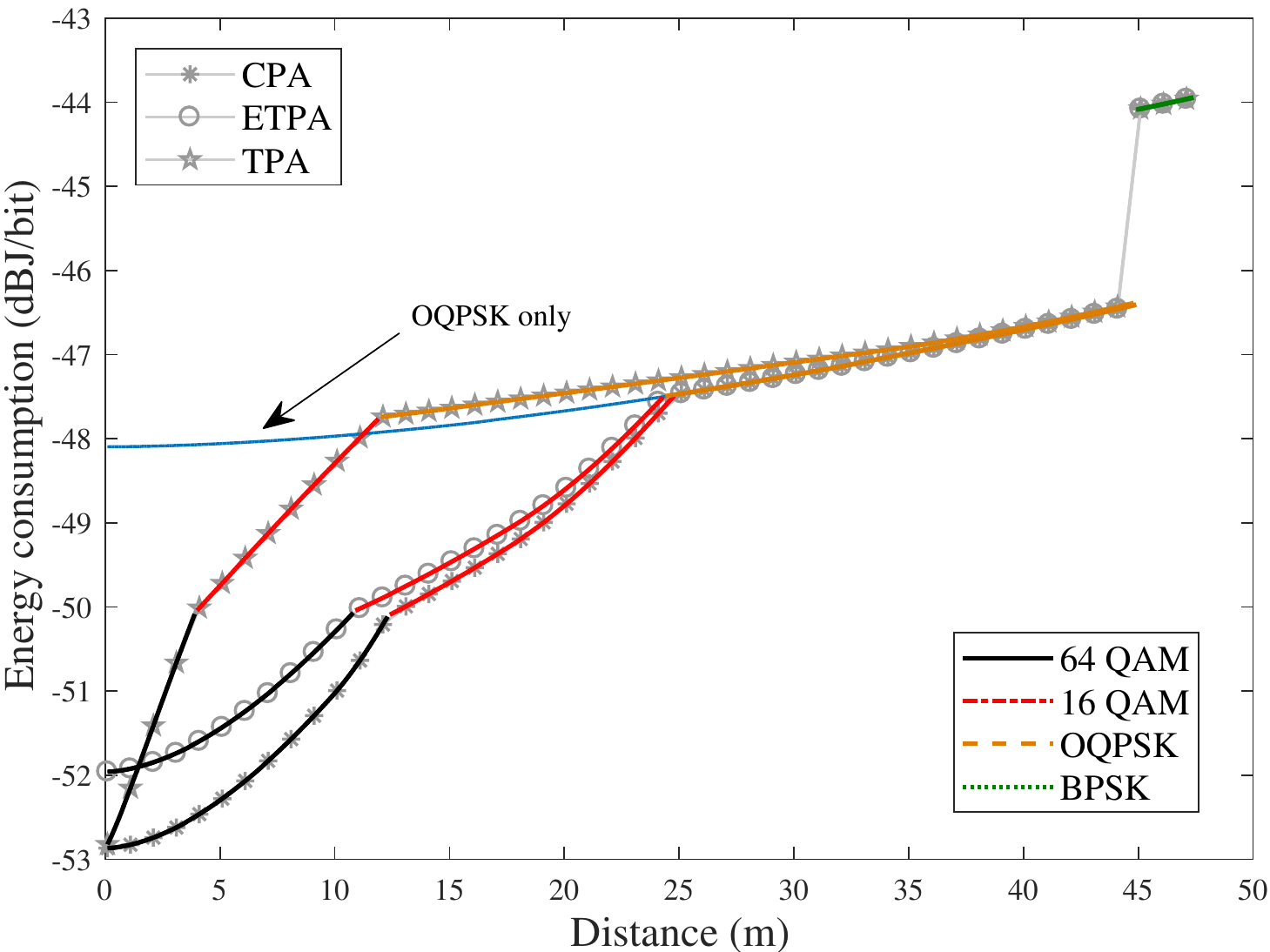}\label{fig:1a}}\hfill
  \subfloat[Optimal transmit power ($P_t$) and PA power consumption ($P_{\textrm{pa}}$)]{%
        \includegraphics[width=0.5\linewidth]{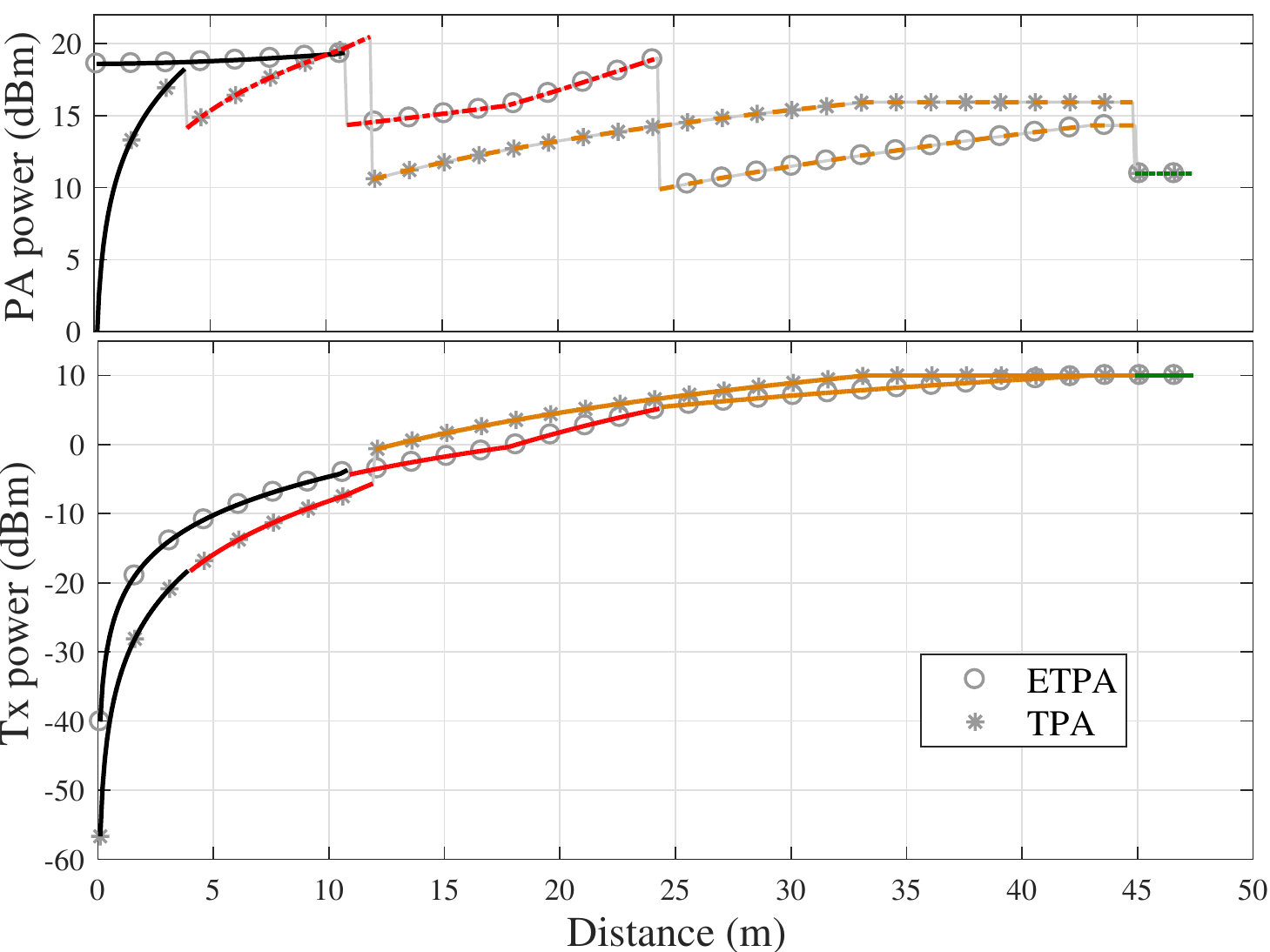}\label{fig:1b}}\\
	\subfloat[Optimal SNR ($\bar{\gamma}$)]{%
        \includegraphics[width=0.33\linewidth]{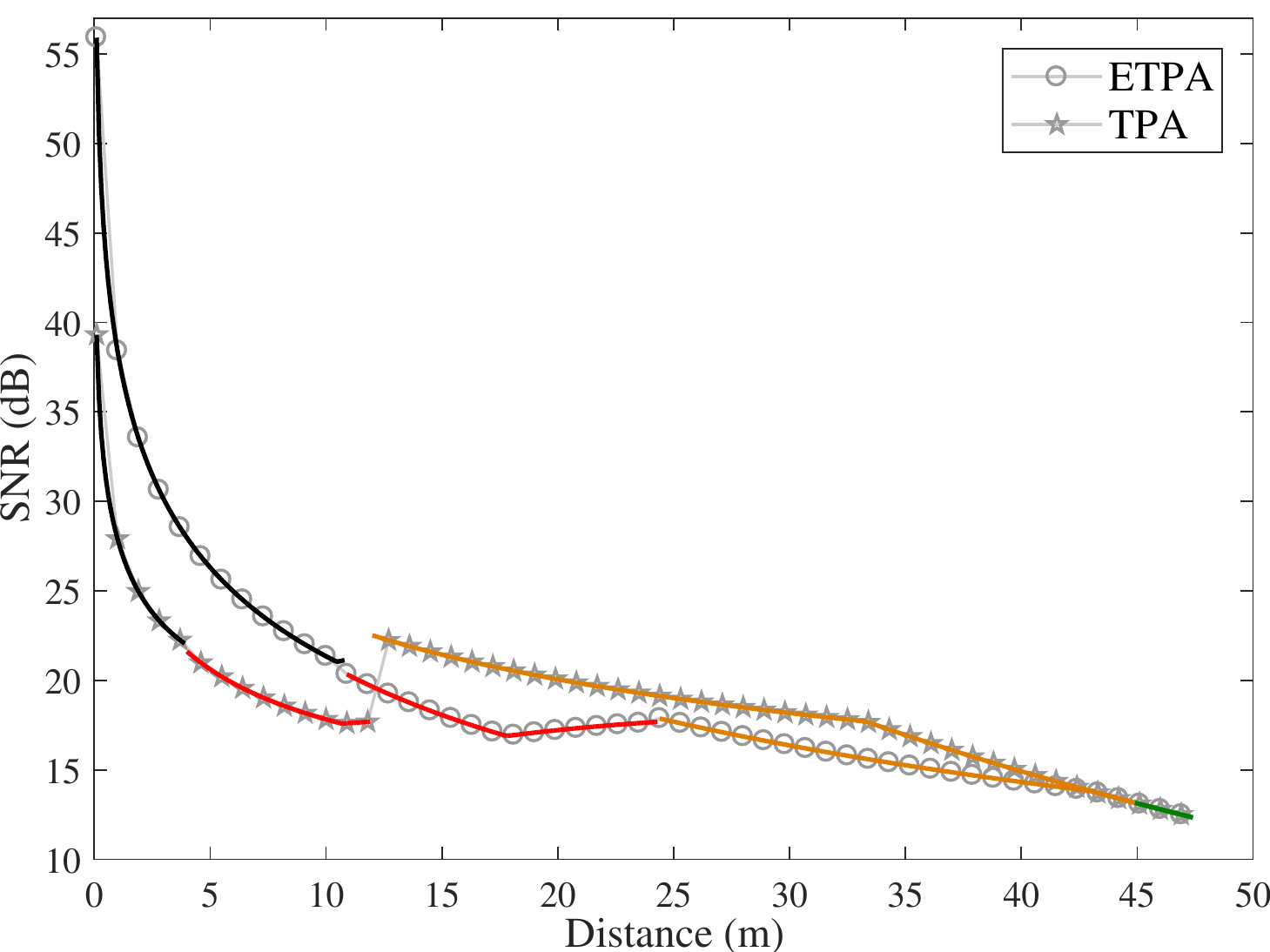}\label{fig:1c}}\hfill
	\subfloat[Optimal payload ($n_p$)]{%
        \includegraphics[width=0.33\linewidth]{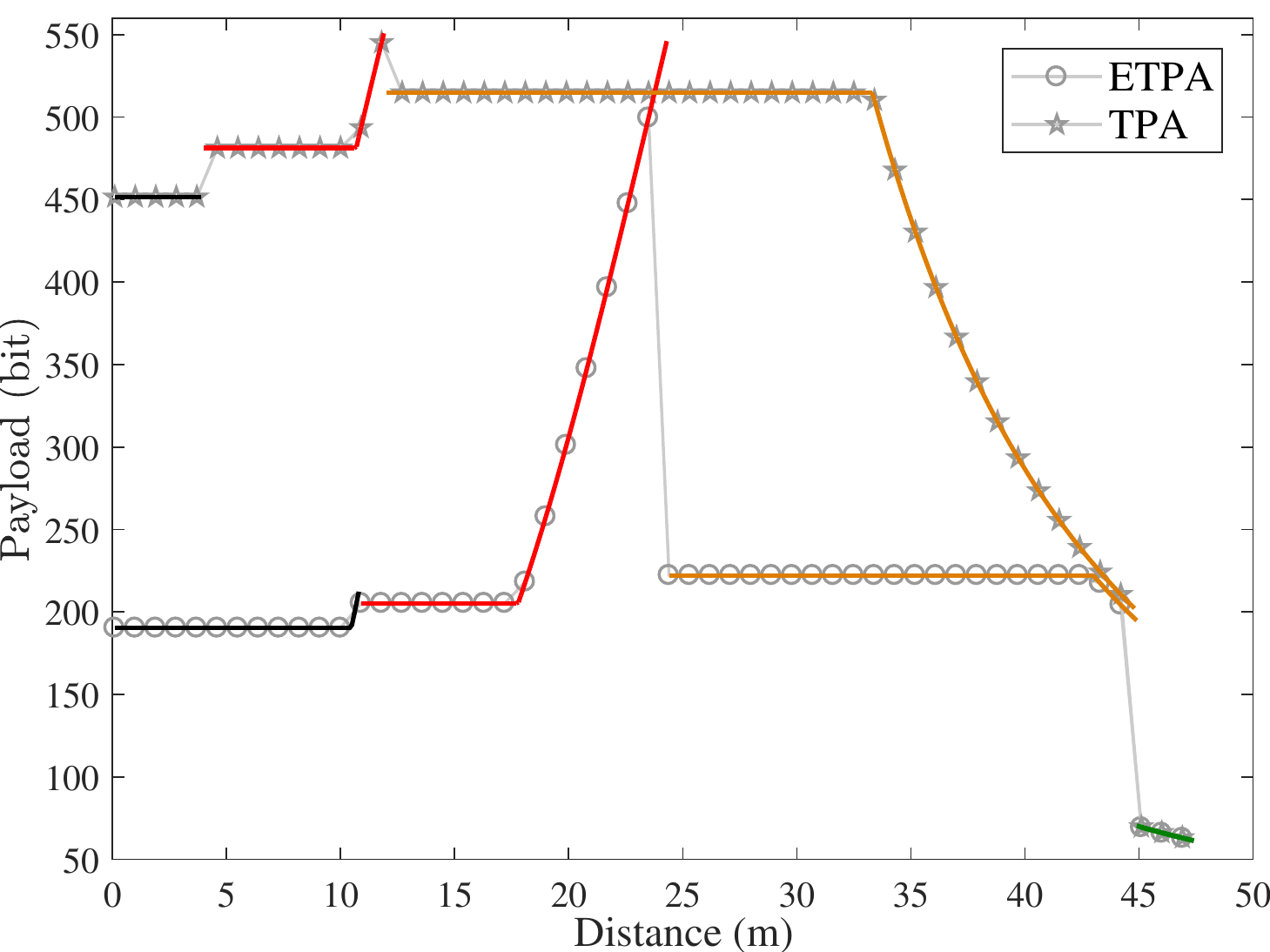}\label{fig:1d}}\hfill
	\subfloat[Optimal no. of retransmissions ($\tau_r$)]{%
        \includegraphics[width=0.33\linewidth]{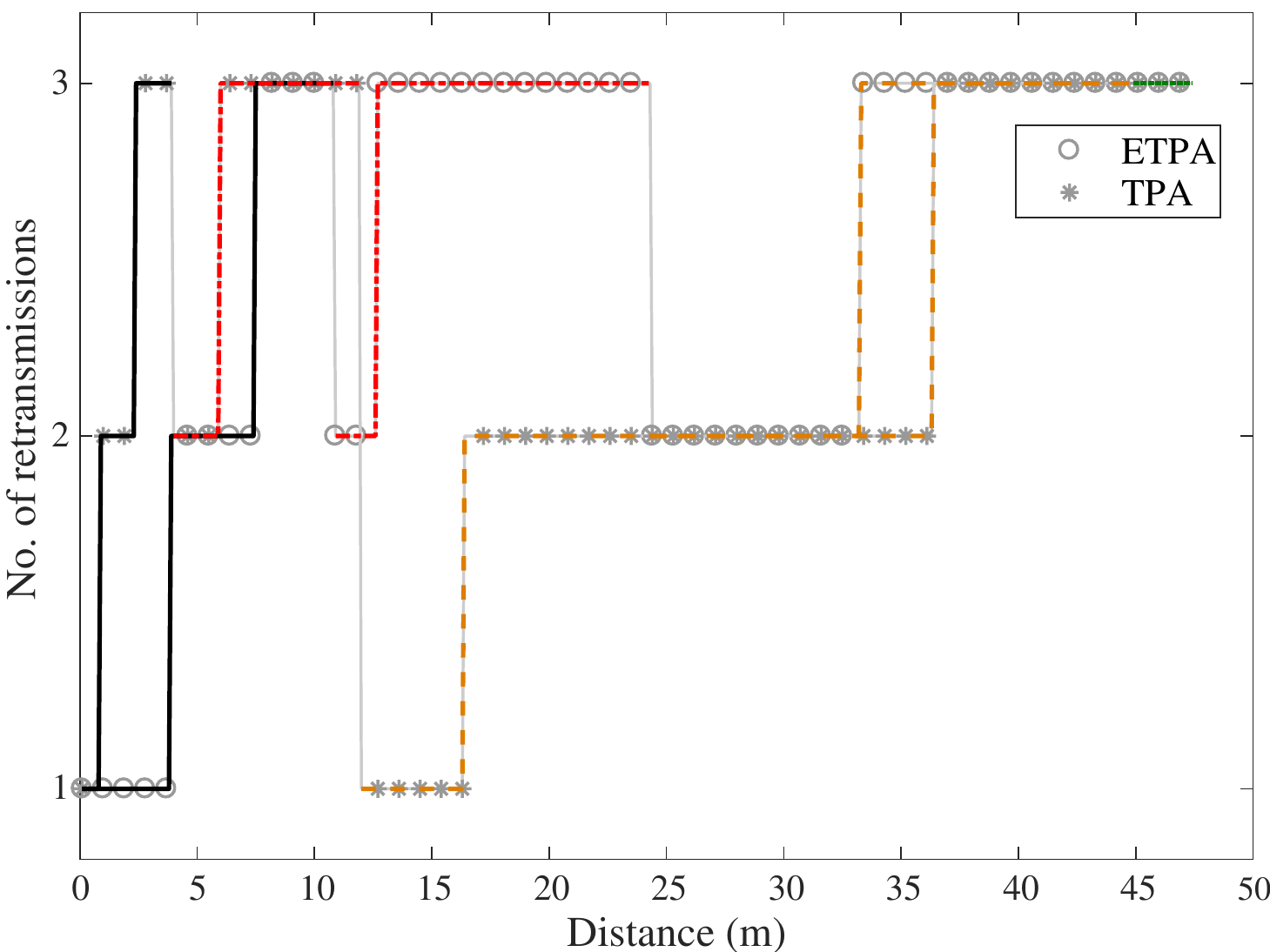}\label{fig:1e}}
  \caption{Impact of PA models on energy consumption and optimal PHY/MAC parameters. 
	Figures (a) and (b) are essentially reciprocal results for 
	$\bar{\gamma}$ and $P_t$, respectively, while (b) shows additional 
	information on PA power consumption ($P_{\textrm{pa}}$).}
  \label{fig:AlgoResult} 
	\vspace{-10pt}
\end{figure*}

Using Algorithm~1, Fig.~\ref{fig:AlgoResult} gives {a} holistic view of 
how energy-optimal parameters---modulation size ($M$), 
SNR ($\bar{\gamma}$) or transmit power ($P_t$), payload ($n_p$) and the 
number of retransmissions ($ \tau_{r}$)---vary with the link distance. 
Mainly, it compares the impact of PA models on the EC while 
closely looking at the optimal system parameters. 
Fig.~\ref{fig:AlgoResult}\subref{fig:1a} shows that the EC 
under constant PA (CPA)-model is considerably optimistic compared to 
realistic PA models. Although the EC for envelope-tracking PA (ETPA) is 
higher and closely follows the EC for CPA, it is considerably higher for 
traditional PA (TPA). As a result, TPA model sees a link switching to a 
low-order modulation at shorter distances. Thus, TPA reduces the gain in 
using high-order modulation for a device using only OQPSK 
modulation.% irrespective of its distance from the controller. 

In Fig.~\ref{fig:AlgoResult}\subref{fig:1b}--Fig. 
\ref{fig:AlgoResult}\subref{fig:1d}, we can observe the trade-off or 
inter-play among the link parameters in minimizing EC under ETPA and TPA, 
where the parameters under CPA are omitted for the clarity of figures. 
Nevertheless, their trend is similar to the link parameters for ETPA. In general, 
the behavior of these parameters must be interpreted together with the 
PA-efficiency curves in Fig.~\ref{fig:EE_PA_Models}. It can be noticed that, for 
a device at short distances, higher-order modulation and high-power 
transmissions are energy efficient under ETPA. Although ETPA has {a} higher power consumption 
($P_{pa} = \xi P_t/\eta(P_t)$) than TPA at the same distances 
(see Fig.~\ref{fig:AlgoResult}\subref{fig:1b}-(top) for distance between 
\SIrange{1}{11}{\meter}), thanks to better linearity of 
ETPA, high-power transmissions are capitalized by the device to {successfully 
transfer} (small-sized) information bits with a low number of 
retransmissions. Whereas, {the} power consumption of TPA is smaller but, 
with the increase in distance, it causes the increase in EC at a higher rate than 
the ETPA. As a result, TPA leads to almost third modulation degradation 
(i.e., to OQPSK) while 64~QAM is still energy optimal under ETPA.
The reason behind low energy efficiency of TPA can be explained from its  
inefficiency at low output powers. To compensate for {the} limited gain {in}  
transmitting at high-power at short distances, the device opt for small 
transmit power but with a higher payload size and {a} higher number of 
retransmissions.

In general, the order of optimal modulation reduces as the distance increases. 
However with ETPA, at distances around \SIrange{11}{24}{\meter}, 16~QAM 
achieves better energy efficiency than TPA owing to its better trade-off in 
efficiency and the other link parameters (as discussed earlier). An interesting 
observation in this range is the rapid increase in payload size with the 
distance, which starts at around \SI{17}{\meter}. However, note that, the 
corresponding SNR (in Fig.~\ref{fig:AlgoResult}\subref{fig:1c}) remains 
constant in this distance range. That is because increasing the transmit power 
with increasing packet size is optimal until next smaller modulation becomes 
energy optimal. If observed carefully, a small jump in payload size before 
the ETPA link downgrades its modulation from 64~QAM to 16~QAM can also be 
noticed.

When comparing the results where the same modulation scheme is employed 
under both the ETPA and TPA (i.e., at distances $\geq$\SI{24}{\meter}), it 
can be observed that both transmit power and power consumption for TPA jumps 
higher than ETPA for the first time and their corresponding EC become almost 
identical. This is because the TPA efficiency increase exponentially in this 
range of output power, and also the small PAPR of OQPSK modulation reduces 
the amount of {back off} from the saturation region. As {a} consequence, this has an 
effect on the link parameters under TPA where the payload starts reducing 
with the distance and the number of retransmissions also reduce at some 
distances. However, as soon as the link operates at its maximum allowed 
transmit power, the link parameters under two PAs, including the modulation 
scheme (i.e., BPSK), become identical.

%%%%%%%%%%%%%%%%%%%%%%%%%%%%%%%%%%%%%%%%%%%%%%%%%%%%%%%%%%%%%%%%%%%%%%%%%%%%%%
%%%%%%%%%%%%%%%%%%%%%%%%%%%%%%%%%%%%%%%%%%%%%%%%%%%%%%%%%%%%%%%%%%%%%%%%%%%%%%
%%%%%%%%%%%%%%%%%%%%%%%%%%%%%%%%%%%%%%%%%%%%%%%%%%%%%%%%%%%%%%%%%%%%%%%%%%%%%%
%%%%%%%%%%%%%%%%%%%%%%%%%%%%%%%%%%%%%%%%%%%%%%%%%%%%%%%%%%%%%%%%%%%%%%%%%%%%%%
%%%%%%%%%%%%%%%%%%%%%%%%%%%%%%%%%%%%%%%%%%%%%%%%%%%%%%%%%%%%%%%%%%%%%%%%%%%%%%
%%%%%%%%%%%%%%%%%%%%%%%%%%%%%%%%%%%%%%%%%%%%%%%%%%%%%%%%%%%%%%%%%%%%%%%%%%%%%%

\subsection{Lifetime Analysis of IoT Links/Devices}
The optimal link parameters, as discussed earlier, allow analyzing the 
lifetime of IoT devices while considering their non-ideal hardware 
characteristics. For the analysis, we assume that each device is 
battery-powered, and it reserves a charge capacity of \SI{2}{\ampere}\SI{}{h} 
only for data communication. Also, each device, located at a certain distance 
from the gateway, {is} assumed to be transmitting \SI{5}{\kilo\bit} of sensory data on the average 
within a period of \SI{5}{\minute}. 
 
\begin{figure}[!t]
	\centering
		\includegraphics[width=0.90\linewidth]{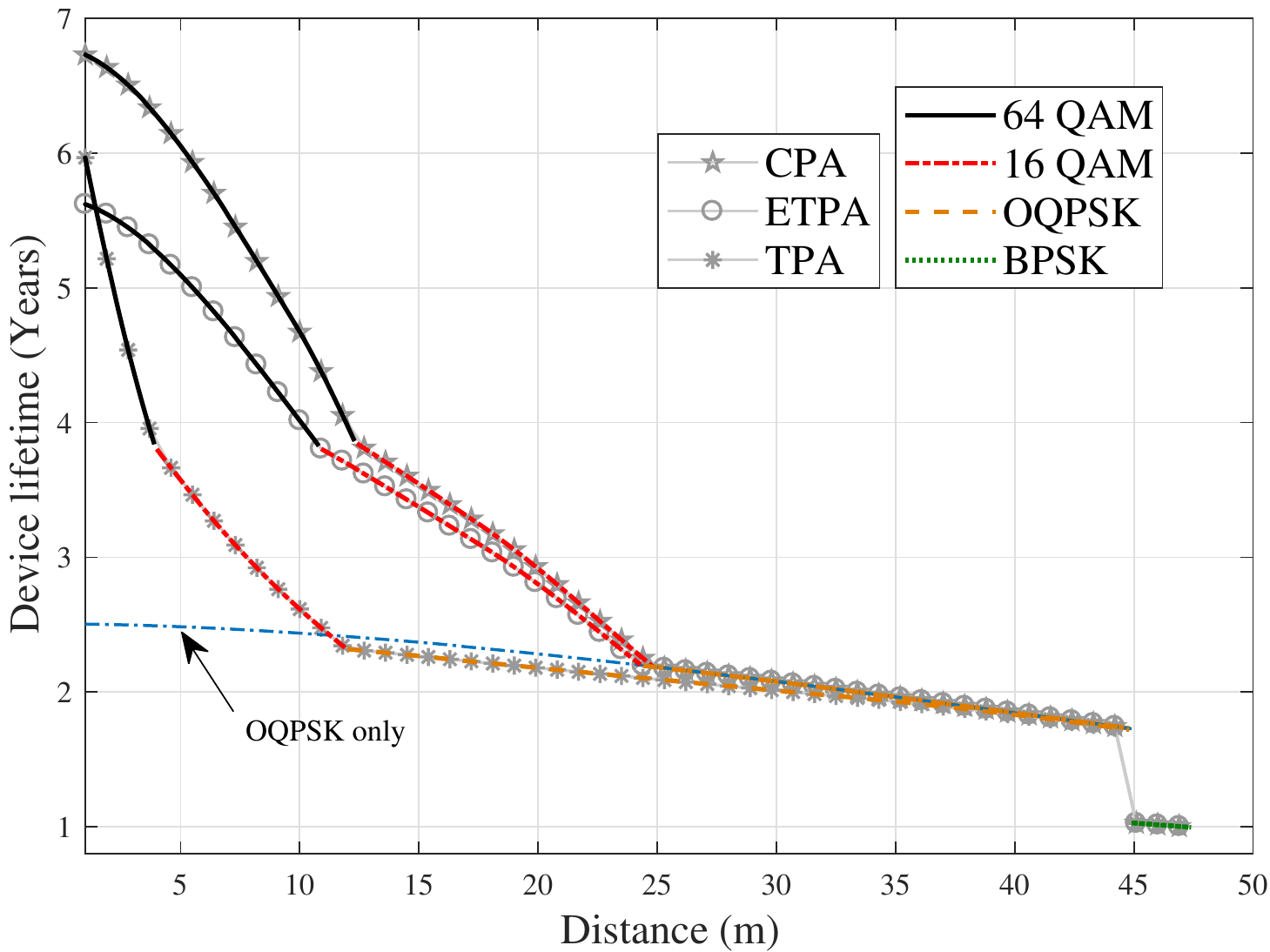}
	%\vspace{-18pt}
	\caption{Lifetime analysis of a device transmitting on average \SI{5}{
\kilo\bit} of sensor data every \SI{5}{\minute}. Reliability parameters: $
\mathcal{P}=\num{0.001}$, $\tau_r^{\max}=\num{3}$, Other parameters: $P_0=\SI{
10}{\milli\watt}$, $n_h=\SI{48}{\bit}$, $\eta_{\max}=\SI{80}{\percent}$.}
	\label{fig:lifetime}
\end{figure}

Fig.~\ref{fig:lifetime} shows the lifetime of a device with respect to link 
distance for the studied PA models, when operated with {the} optimized link 
parameters. It also depicts the lifetime of a device 
using only OQPSK modulation while the selection of other link parameters 
at any distance is energy-reliability optimal. We observe that spectral 
efficient modulations can significantly prolong the lifetime of the devices 
located at short-range distances, i.e., within \SIrange{1}{24}{\meter} under 
CPA and ETPA. The expected lifetime increases as the distance decreases, and 
compared to a commonly used low-order modulation (i.e., OQPSK) in IoT 
devices, the lifetime can be extended up to \SI{180}{\percent} in case of an 
ideal PA and \SI{125}{\percent} for ETPA. On the other hand, traditional PA 
(TPA) not only halves the range in which a high-order modulation can help in 
extending the device/link life, it also brings down the gain in using 
higher-order modulations significantly.   

%%%%%%%%%%%%%%%%%%%%%%%%%%%%%%%%%%%%%%%%%%%%%%%%%%%%%%%%%%%%%%%%%%%%%%%%%%%
%%%%% Conclusions
%%%%%%%%%%%%%%%%%%%%%%%%%%%%%%%%%%%%%%%%%%%%%%%%%%%%%%%%%%%%%%%%%%%%%%%%%%%
%\vspace{-8pt}
\section{Conclusions} 
\label{sec:Conclusions}
We studied cross-layer optimization of battery-operated wireless links with 
energy consumption (EC) minimization objective while considering: (a) 
reliability requirements of IoT applications, and (b) operational constraints 
and non-ideal characteristics of the IoT devices' hardware. To this end, we derived EC 
models
%---giving energy cost for delivering an information bit without 
%error---
while capturing energy cost of ideal and realistic power amplifiers
(PAs), and packet error statistics in particular. For packet error statistics, we developed an accurate PER 
approximation in Rayleigh block-fading, which is simple to exploit for 
cross-layer link design and optimization. Using the EC models, we derived 
energy-optimal, reliability-aware, and hardware-compliant conditions for 
SNR and payload size, which we exploited for developing a holistic 
algorithm to optimize the link parameters jointly. The algorithm can be utilized 
by resource-constrained devices for link adaptation based on the optimal 
selection of parameters.% with respect to distance.

Our path to link optimization allowed to make useful observations, especially 
when the target is to simultaneously minimize energy consumption and ensure certain 
reliability. We found that at each distance there is an 
optimal SNR and payload size, which leads to an energy optimal modulation 
scheme where the modulation order increases for short-range links. However a 
reliability-aware link compared to a delay tolerant system reduces this 
distance, and increases the energy 
consumption. Also, when non-realistic power amplifier (PA) characteristics are 
considered, we found that the traditional PA significantly diminishes 
the gain in the usage of higher-order modulations (causing both the higher energy 
consumption for higher-order modulations, and switching to smaller 
modulations at shorter distances). While an envelop-tracking PA behaves 
closely to an ideal PA. Since the higher-order modulations offer packing more 
information bits, they are energy-optimal at short distances, because circuit 
power exceeds PA power consumption. However, PA's non-linearity makes these 
powers comparable and leads to a higher energy consumption compared to an 
ideal PA.

Our lifetime analysis found that under ideal PA the lifetime can be 
extended up to \SI{180}{\percent} by selecting higher-order modulations in 
short-range networks compared to OQPSK. However, this gain reduces 
significantly 
under traditional PA, and the distance up to which a higher-order modulation 
can provide any gain reduces to half. Whereas, the envelop-tracking PA 
can still provide the lifetime extension of up to \SI{125}{\percent}. These 
results highlight the need to investigate efficient PAs for 
short-range IoT networks in order to gain from spectral efficient modulations. 

%{In future, we would like to study link 
%optimization under the impact of non-ideal PAs in interference and multihop communication 
%scenarios.}

{As a future work, the presented link optimization under 
non-linear PAs can be refined for relay selection and transmission 
duty-cycle optimization in an IoT system, as in~\cite{tan2018cross}.}

%\vspace{-2pt}
\appendix[Proof of Proposition 1]
For an $N$-bit packet, the PER function in 
\eqref{eq:p_awgn1} for BER functions described by $c_m e^{-k_m\gamma}$ and 
$c_m Q(\sqrt{k_m\gamma})$ is asymptotically approximated by the Gumbel 
distribution function for the sample minimum \cite{Our2016}
\begin{equation}
f(\gamma) \simeq 1-\mathrm{exp}\left(-\mathrm{exp}\left(-\frac{\gamma-a_N}{b_N
}\right)\right),
\label{eq:gumbelDistributionApprox}
\end{equation}
where $a_N$ and $b_N>0$ are the normalizing constants.

Let $G\left(\gamma\right) = \mathrm{exp}(-\mathrm{exp}(-
\frac{\gamma-a_N}{b_N}))$ be the cumulative distribution function (CDF) of 
the Gumbel distribution for the sample maximum, then from 
\eqref{eq:gumbelDistributionApprox} and \eqref{eq:threshold_0} we have
\begin{equation}
\omega_0 \approx \int_0^\infty {\left(1-G\left(\gamma\right)\right)d\gamma}, 
\label{eq:threshold_new}
\end{equation}
which is the expected value of a continuous and non-negative random variable $
\gamma$.

%\balance
\bibliographystyle{IEEEtran}
\bibliography{bib}

\end{document}